\documentclass[12pt,a4paper]{iopart}
\pdfoutput=1

\usepackage{caption}
\usepackage{multirow}
\usepackage{iopams}  
\usepackage[utf8]{inputenc}
\usepackage{listings}
\usepackage[ruled,vlined]{algorithm2e}
\usepackage{float}
\usepackage{graphicx}
\usepackage{subfigure}
\usepackage{tikz}
\usepackage{adjustbox}
\usepackage{xcolor,colortbl}
\usepackage{lipsum}
\usepackage{epsfig}
\usepackage{epstopdf}
\usepackage{cuted}

\usepackage{chngcntr}

\tikzset{
  int/.style={circle, draw, fill=blue!20, minimum size=3em},
  init/.style={pin distance=1.2cm,pin edge={loop,thin,black}}
}
\usetikzlibrary{arrows}
\expandafter\let\csname equation*\endcsname\relax
\expandafter\let\csname endequation*\endcsname\relax
\usepackage{amsfonts}
\usepackage{amsmath,amsthm,mathtools}
\usepackage{lineno,hyperref}
\modulolinenumbers[5]

\DeclareCaptionFormat{cont}{#1 (cont.)#2#3\par}

\begin{document}

\title[M-Centrality]{M-Centrality: Identifying key nodes based on global position and local degree variation}

\author{Ahmed IBNOULOUAFI$^1$$^,$$^{a}$, Mohamed EL HAZITI$^2$,
Hocine CHERIFI$^3$}
\address{$^1$ LRIT Laboratory, Associated Unit to CNRST (URAC29) IT Rabat Center -  Faculty of Sciences In Rabat, MOHAMMED V UNIVERSITY IN RABAT, B.P.1014 RP, Rabat, Morocco}
\address{$^2$ Higher School of technology (E.S.T) in SALE}
\address{$^3$ Laboratoire Electronique, Informatique et Image (Le2i) UMR 6306 CNRS, University of Burgundy, Dijon, France.}
$^{a}$ Corresponding author: \ead{ahmedibnoulouafi@gmail.com}

\begin{abstract}
Identifying  influential nodes in a network is a major issue due to the great deal of applications concerned, such as disease spreading and rumor dynamics. That is why, a plethora of centrality measures has emerged over the years in order to rank nodes according to their topological importance in the network. Local metrics such as degree centrality make use of a very limited information and are easy to compute. Global metrics such as betweenness centrality exploit the information of the whole network structure at the cost of a very high computational complexity. Recent works have shown that combining multiple metrics is a promising strategy to quantify the node's influential ability. Our work is in this line. In this paper, we introduce a multi-attributes centrality measure called M-Centrality that combines  the information  on the position of the node in the network with the local information on its nearest neighborhood. The position is measured by the K-shell decomposition, and the degree variation in the neighborhood of the node quantifies the influence of the local context. In order to examine the performances of the proposed measure, we conduct experiments on small and large scale real-world networks from the perspectives of transmission dynamics and network connectivity. According to the empirical results, the M-Centrality outperforms its alternatives in identifying both influential spreaders and nodes essential to maintain the network connectivity. In addition, its low computational complexity makes it easily applied to large scale networks.
\end{abstract}

%
\vspace{2pc}

\maketitle

\section{Introduction}

The problem of identifying nodes that are ''central'' or ''influential'' has attracted wide attention from researchers due to its many applications. In epidemic spreading, the transmission of the disease depends on the contacts that the infected person has with the susceptible population. Thus, being able to locate and vaccinate the most influential individuals can prevent from a potential outbreak of the disease \cite{lee2012exploiting}. In viral marketing, being able to locate influential individuals can help to optimize the sales of products \cite{leskovec2007dynamics}.

A straightforward approach towards detecting these central nodes is to quantify their influence using centrality measures. The idea of centrality was initially introduced in the context of sociology to look whether there is a  relation between the location of an individual in the network and its influence in group processes. Since then, various centrality measures have emerged over the years. They are employed in a multitude of contexts to rank nodes according to their topological importance. As there is no consensual definition of the centrality of a node within a network, the issue is considered  from the multiple concepts reflecting the notion of influence. Therefore, we can classify centralities from different perspectives. They can be classified according to the underlying approach (geometric, spectral or path-based) \cite{boldi2014axioms}, the way they quantify influence (locally, globally, position within the network, dynamical processes such as random walks)  \cite{lu2011leaders, li2014identifying}, the computing ideas (iterative refinement) \cite{ren2014iterative}, the number of attributes they take into consideration \cite{ren2014iterative} or by taking into account the collective influence of the whole set of nodes, i.e., in this case, the problem of finding influential spreaders is targeted as an influence maximization problem \cite{kempe2003maximizing,morone2015influence,he2016correction}, where we look for a minimum set of nodes that maximize the spread of information to the whole network.

In geometric measures, influence is related to distances. In other words, it depends only on how many nodes exist at every distance. Degree \cite{bonacich1972factoring}, closeness \cite{du2015new} and K-shell \cite{kitsak2010identification} (and its variations, including MDD \cite{zeng2013ranking} and INK \cite{lin2014identifying}) are geometric measures that evaluate influence from respectively a local, global and position point of view. In \cite{chen2013identifying}, authors proposed a local ranking measure, ClusterRank, that quantifies the influence of a node by taking into account not only its direct influence (measured by the number of its followers) and influences of its neighbors, but also its clustering coefficient. ClusterRank can  be applied to directed as well as undirected networks where its superiority in term of locating influential spreaders is significant compared with degree centrality, K-shell decomposition,  PageRank and LeaderRank. In addition, ClusterRank, only making use of local information, is much more efficient considering computational complexity. Spectral measures work on the spectral properties of the graph (eigenvalues and eigenvectors of the adjacency matrix). Iterative refinement methods identify influential nodes based on the influence of its neighbors. This is known as the mutual enhancement effect.  PageRank \cite{grin1998anatomy}, HITS \cite{kleinberg1999authoritative}, LeaderRank \cite{lu2011leaders} and Personalized PageRank \cite{nathan2017local} (where the probability of jumping to a node when abandoning the random walk is not uniform, but it is given by a preference vector that favorites nodes over others) are good examples of these methods. Path-based measures exploit the existence of shortest paths passing through a node.  Betweenness \cite{prountzos2013betweenness} is the best known centrality of this type. It evaluates influence from a global perspective by considering influential nodes those through which transits the largest flow of information. As an alternative to betweenness centrality, a centrality measure called DIL, was proposed in \cite{liu2016evaluating}. Instead of working on a global level, DIL centrality ranks nodes based on local information (degree value and the importance of lines) to identify network bridges. This measure showed great performances and is adapted to large scale network due to its low complexity. Note that global and spectral measures are inapplicable to very large graphs due to their computational complexity, while local ones are simple but generally less effective because they only take in consideration the neighborhood of nodes. Another way of approaching the problem of locating influential nodes is to optimize an objective function of influence. In this context, Morone and Makse proposed a method called Collective Influence expressed as the product of the reduced degree of a node and the total reduced degree of all nodes at a distance d from the node \cite{morone2015influence}. According to their results, optimal results can be reached at a distance d=3. The Collective Influence formula constitutes the main core of their percolation algorithm used to find the minimal set of nodes which are crucial for the global connectivity of the network (their removal cause the destruction of the giant component). The major drawback of this method is its computational complexity that stands at $\mathcal{O}(nlog(n))$ \cite{morone2015influence,lu2016vital}.

Recent works have raised the fact that centrality cannot be apprehended from a single point of view and that combining measures can enhance the performances of ranking methods \cite{gao2015combination}. Comprehensive evidence centrality (CEC) \cite{mo2015evidential} is a multi-attributes method that uses the Dempster-Shafer evidence theory to combine degree, betweenness and closeness centralities in order to characterize the influence of a node. Another multi-criteria method proposed in \cite{liu2015node} uses the Technique for Order Preference by Similarity to Ideal Solution (TOPSIS) to evaluate the influence of each node from more than one perspective based on four indicators which are the degree, betweenness, closeness and improved K-shell (IKs). Dynamic sensitive centrality (DS) \cite{liu2016locating} is also a multi-attributes method that takes in consideration topological features (a dynamical process represented by the calculation of the total number of walks of length t from node i to all nodes in the network) and dynamical properties that depend on the spreading rate $\beta$ over the time. DS centrality can locate influential nodes accurately and performs very well in the early stages of spreading compared to degree, K-shell and eigenvector centralities. In \cite{ma2016identifying}, Ma et al. proposed a centrality (labeled Gravity) based on the Isaac Newton classical gravity formula. Gravity centrality considers the K-core value of a node as its mass, and the shortest path distance between two nodes in a network is viewed as their distance. To reduce the complexity of their method, only neighbors located in a distance less than or equal to 3 are considered. This method shows that combining a position based measure (Coreness) and geodesic distance, gives better results in term of locating influential spreaders. 

The main lesson learned from these studies is that combining multiple attributes is much more accurate than using a single one in order to evaluate the influence capability of a node. However, this raises new questions such as which centralities should be combined and  how to combine them.  Previous studies show that position \cite{kitsak2010identification} and neighborhood \cite{zhong2015iterative,xu2017iterative} are key factors to quantify the node influence. Based on this evidence, we propose a new centrality measure combining these complementary aspects.  The measure is called M-Centrality in homage to the M-Theory that unifies all consistent versions of the super-string theory. It is a weighted combination of :
\begin{itemize}
\item A "global" measure that characterizes the location importance of the node in the all network (core or periphery). We use the K-shell decomposition for its low computational complexity. Note, however, that any of its variations can be substituted.
\item A local measure that  characterizes its neighborhood. This measure we choose to call $\Delta\mathfrak{D}$ calculates the degree variation in the neighborhood of the node.  It is inspired from the preferential attachment process \cite{zhao2011analyzing}. 
\end{itemize}
Since it cannot be assumed that attributes have equal weights, the Shannon entropy method  is used  in order to find the appropriate weights for  the global and local measure. It  is one of the most famous approach for determining the objective attribute weights
in multiple-criteria decision problems.

To evaluate the proposed centrality measure, we report a series of experiments on real-world networks. Extensive comparisons with the most influential alternative measures are performed. Results clearly show that globally, M-Centrality  provides more accurate ranking list. In addition to its effectiveness, one main advantage of this measure as compared to alternative global ranking methods is that it has a low computational complexity, i.e. $O(n)$ (as shown in Section \ref{regsol0}), that allows it to be used with large scale networks.

The remainder of this article is organized as follows. In section 2, we review the necessary background on evaluation metrics used to asses different centralities and the preferential attachment concept that inspired our approach. Section 3 introduces the M-Centrality measure, and it shows how it is related to the preferential attachment process.  In section 4, the datasets, the experimental setup and experimental results are presented. The proposed measure is analyzed and compared to the most recent methods, including Gravity (Gr), DIL, Personalized PageRank (PPR), ClusterRank (CR) and Collective influence (COI). Finally, section 5 concludes the paper.

\section{Background}
In this section, we recall the definition of the evaluation measures that are used to compare centralities (Monotonicity, Kendall's Tau correlation coefficient ($\tau$) and decline rate of network efficiency). Additionally, the model used to simulate an epidemic spreading process in order to evaluate the performances of the different methods in the context of transmission dynamics is presented. Finally, we present the preferential attachment process. 

\subsection{Evaluation of nodes ranking methods}

The main role of centrality measures is to provide a means to rank nodes relative to each other. Indeed, the numerical values may not be directly interpretable. Nodes are usually ranked in descending order of influence. The node with largest influence is  ranked first and the one with smallest influence value is ranked last. There are many ways to evaluate the performances of centrality measures, more or less linked to the underlying applications. In this paper, we adopt generic measures (Monotonicity and Kendall's Tau correlation coefficient) that are commonly employed in the literature to quantify influence.  Monotonicity measures the ability of a ranking method to assign a different rank to each node, while Kendall tau correlation allows a statistical comparison of the agreement between two rankings. In addition, we also evaluate the performances of centrality measures from two perspectives, one is based on transmission dynamics where we look for the spreading capabilities of nodes, and the other is based on the network connectivity and the theory that the network damage caused by deleting a node is equivalent to its importance.

\textit{Monotonicity:}\\
To quantify the resolution of a ranking method, the monotonicity $\mathfrak{M}$ \cite{bae2014identifying} of a ranking vector $R$  is defined as follows:

\begin{equation}
\mathfrak{M}(R)= \Bigg(1 - \frac{\sum_{r \in R}n_r(n_r - 1)}{n(n - 1)}\Bigg)^2,  \mathfrak{M} \in [0, 1]
\label{eq:equation18}
\end{equation}

where $n$  is the size of ranking vector R and $n_r$ is the number of ties with the same rank $r$ . This metric quantifies the fraction of ties in the ranking list. The monotonicity $\mathfrak{M}$(R) is equal to one if the ranking vector R is perfectly monotonic, and it is equal to zero if all nodes in R have the same rank. Note that monotonicity allows to quantify the discrimination ability of a centrality measure, however, a ranking with no tie is not necessarily accurate.

\textit{Kendall tau correlation:}\\
The Kendall's Tau ($\tau$) correlation coefficient \cite{nelsen2001kendall} is generally used to compare the performance of different topology-based measures. It measures the ranking consistency of two lists that rank the same set of objects. Consider X = ($x_1$, $x_2$, ..., $x_n$) and Y= ($y_1$, $y_2$, ..., $y_n$) are two ranked lists that contain n elements, respectively. Any pair of ranks ($x_i$, $y_i$) and ($x_j$, $y_j$) is said concordant if $x_i$ $>$ $x_j$ and $y_i$ $>$ $y_j$ , or if $x_i$ $<$ $x_j$ and $y_i$ $<$ $y_j$ . If $x_i$ $>$ $x_j$ and $y_i$ $<$ $y_j$, or if $x_i$ $<$ $x_j$ and $y_i$ $>$ $y_j$, then the pair is said to be discordant. In the case of $x_{i}=x_{j}$ or $y_{i}=y_{j}$ (tied pair), the pair is neither concordant nor discordant, in this case  X and Y are independent.  The Kendall's Tau ($\tau$) correlation coefficient is defined by: 
\begin{equation}
\tau_{X,Y}= \frac{n_c - n_d}{\sqrt{(n_0 - n_1)(n_0 - n_2)}}, \tau \in [-1, 1]
\label{eq:equation14}
\end{equation}
where $n_0 = n(n - 1)/2$, $n_1 = \sum_{i}t_i(t_i - 1)/2$, $n_2 = \sum_{j}t_j(t_j- 1)/2$, $n_c$ and $n_d$ are respectively the number of concordant pairs and discordant pairs, $t_i$ and $t_j$ are the number of tied values in the $i^{th}$ and $j^{th}$ group of ties respectively.  We can consider the following ranges in order to qualify the strength of the relation between two rankings:

\begin{itemize}
\item No correlation: $\tau$ = 0, 
\item Low correlation: $\tau$ $\in$ $]$0, 0.50$[$, 
\item Moderate correlation: $\tau$ $\in$ $[$0.50, 0.70$[$, 
\item High correlation: $\tau$ $\in$ $[$0.70, 0.90$[$,
\item Very high correlation: $\tau$ $\in$ $[$0.90, 1$[$,
 \item Perfect correlation: $\tau$ = 1.
\end{itemize}

\textit{SIR model:}\\
In literature, the susceptible-infected (SI) and susceptible–infected–removed (SIR) models  \cite{sutton2014discretizing,1742-5468-2012-07-P07005} are generally used to simulate an epidemic spreading process in real networks. Compared to the SI model, the SIR model is widely used for information dissemination and disease diffusion and in various fields. In this paper, we employ the SIR model to estimate the spreading capabilities of the nodes. In this model, a node has three possible states: S (susceptible), I (infected) and R (recovered). Initially we start from a single infected node and the other nodes are susceptible. At each step, the infected node can infect its susceptible neighbors with infection probability $\beta$, and then it recovered from the diseases with probability $\gamma$ (set to 1 in this paper). The spreading process stops when there is no infected node in the network. At last, the number of recovered nodes represents the influence of the node. In this article, for the SIR epidemic model we consider:

\begin{itemize}
\item For each node we simulate $10^2$ diffusion process. The spreading capability of the
node will be the average number of infected nodes.
\item We chose values of infection rate $\beta$ that vary from $20\%$ to $160\%$ of $\beta_{th}$ (the epidemic threshold \cite{pastor2002immunization}) representing both cases where $\beta$ $<$ $\beta_{th}$ and $\beta$ $\geq$ $\beta_{th}$.
\end{itemize}

\textit{Network efficiency:}\\
Network efficiency \cite{ren2013node} reflects the network connectivity. The better the network efficiency is, the better the network connectivity is. Network efficiency $\eta$ is defined thus:

\begin{equation}
\eta = \frac{1}{n(n - 1)}\sum_{i \neq j \in V}\eta_{ij},
\end{equation}

where $\eta_{ij}$ is the efficiency between i and j, $\eta_{ij} = \frac{1}{d_{ij}}$, $d_{ij}$ is the shortest way between i and j, n is the number of network nodes.
The decline rate of network efficiency $\nu$ is defined as:

\begin{equation}
\nu = 1 - \frac{\eta}{\eta_0},
\end{equation}

where $\eta$ is the efficiency of the network after removing nodes. $\eta_0$ is the initial efficiency of the network. The bigger the $\nu$ is, the worse the network connectivity destroyed by removing nodes is and the more important the node removed is.

\subsection{Preferential attachment}

Growth and preferential attachment are the two mechanisms used in the most prominent approach to reproduce complex networks formation. Growth is a network construction process where at each time step, a new node with $m$ links is added to the existing network \cite{zhao2011analyzing}. The preferential attachment rule specify that new nodes select old nodes with which they will form links based on their degree. That is, the probability $\Pi_{n\rightarrow i}$ that a new node $n$ makes a connection to an existing node $i$ with degree $k_i$ is given by :

\begin{equation}
\Pi_{n\rightarrow i} (k_{i})= \frac{k_{i}}{\sum_{j \in all} k_{j}},
\label{equation4}
\end{equation}
where $all$ is the set of nodes to which the new node $n$ could connect.\\

Due to this preference, a ''Rich get Richer'' process takes place where the nodes with higher degrees will further increase their connexions leading to the emergence of hubs. An important feature of this model introduced by Barab{\'a}si and Albert \cite{barabasi1999emergence} is that the generated networks display a power-law degree distribution. The growth mechanism combined with preferential attachment is therefore the most influential explanation for the prevalence of the scale-free networks that are ubiquitous in nature and man-made systems.

\section{Proposed method}
\subsection{Motivation}

One of the main application of centrality measures is the assessment of node's spreading capability in the context of epidemic spreading. Various approaches based on centralities have been proposed so far, one of them is based on node degrees \cite{freeman1978centrality}. However, as shown in \cite{bae2014identifying, wang2016fast}, it is not sufficient since it is just the number of neighboring nodes that is considered. An improvement of the approach consists in taking into account the degrees of the neighbors of the node; a node with neighbors that have high degrees has greater spreading capability. It is worth pointing out, however, that the high degree of a node, or its neighbors, is not sufficient for specification of its spreading capability \cite{zareie2017influential}. On the basis of this observation, and driven by the fact that quantifying node influence from multiple points of view, we decided to introduce a new multi-attributes centrality, labeled M-Centrality, that takes in consideration global and local features of a node. Of course there is as much combination possible as the number of centralities, but the main reason behind the choice of K-shell and degree in our combination is their great performances in detecting influential spreaders. Indeed the K-shell shows the best results in the case of single origin spreading \cite{kitsak2010identification} while the degree operates the best in the case of multiple origins spreading \cite{motter2004cascade, basaras2013detecting}.

\subsection{M-Centrality measure}

The main idea of this approach is that both position and neighborhood attributes play important roles in shaping node influence. Therefore, by combining these two attributes, we can enhance the performance of the centrality evaluation process. M-Centrality expresses how influential is a node based on the combination of the local information contained in its neighborhood and a more global information about its position in the network. More precisely, the M-Centrality of the node $i$ that we note $M_{i}$  is the weighted sum of a global measure $K_{s_{i}}$ and a local measure $\Delta\mathfrak{D_{i}}$:
\begin{equation}
 M_{i}= \mu K_{s_{i}} + (1 - \mu) \Delta\mathfrak{D_{i}} , 0 \leq \mu \leq 1
\label{eq:equation9}
\end{equation}
\begin{itemize}
\item $K_{s_{i}}$  is the coreness index. It is computed using K-core decomposition. This global measure characterizes the position (core or periphery) of the node $i$ in the network. Note that any of its variation can be substituted.\\
\item $\Delta\mathfrak{D_{i}}$ is a new measure that we introduce to quantify the degree variation at a local level. It takes inspiration from the preferential attachment process. We first calculate the degree variation ($\mathfrak{d_{i,j}}$) between node $i$ and each of its neighbors as follows:
\begin{equation}
\mathfrak{d_{i,j}}= | N_{i}|.\Big|\frac{k_{j} - k_{i}}{\sum_{j \in N_{i}}k_{j}}\Big|, j \in N_{i}
\label{eq:equation6}
\end{equation}
where $k_{i}$ and $k_{j}$ represent the degree of nodes $i$ and $j$  respectively, $N_{i}$ is the neighborhood of the node $i$ and $| N_{i}|$ the number of $i$ neighbors. We use the absolute value to eliminate negative values in  case of $k_{j}$ $>$ $k_{i}$. 
The degree variation in the neighborhood of node $i$ is given by:
\begin{equation}
\Delta\mathfrak{D_{i}}= \sum_{j}{\mathfrak{d_{i,j}}},
\label{eq:equation8}
\end{equation}
\item $\mu$ is a tailored weighting factor that is estimated from the data. Unlike most traditional multi-attribute ranking methods that consider all attributes as equally important (equal weights),  we propose to compute the weight $\mu$ by targeting the problem in a multi attribute decision-making framework. Among the various solutions, we choose an entropy technique that is known for its great performances in attributes weights determination \cite{he2016information}. The weight computation process proceeds as follows:\\
First, we normalize the global and local measure attributes of the M-Centrality: 

$r_{1j} = \frac{K_{s_{j}}}{\sum_{j=1}^{n}K_{s_{j}}}$ and $r_{2j} = \frac{\Delta\mathfrak{D}_{j}}{\sum_{j=1}^{n} \Delta\mathfrak{D}_{j}}$.\\
 where  $n$ is the size of the network\\
Second, we build the matrix $R_{2,n}$ defined by:
\begin{equation}
R_{2,n}=
\begin{bmatrix}
 r_{11}& r_{12}& r_{13}&\dots&r_{1n}\\
 r_{21}& r_{22}& r_{23}&\dots& r_{2n} 
\end{bmatrix}
\label{eq:equation11}
\end{equation}

Third, we compute the entropy $E_{i}$ of the $i^{th}$ attribute:
\begin{equation}
E_{i}= - \frac{1}{\ln(n)}\sum_{j=1}^{n} r_{ij}\ln(r_{ij}), i =1, 2\ and\ 1 \leq j \leq n
\label{eq:equation12}
\end{equation}
Finally, the weight of the two attributes is computed:
\begin{equation}
w_{i}= \frac{1 - E_{i}}{2 - \sum_{i=1}^{2} E_{i}}, i = 1, 2
\label{eq:equation13}
\end{equation}
According to the properties of entropy, 0 $\leq$ $w_{i}$ $\leq$ 1 and $\sum_{i=1}^{2} w_{i}$ = 1, thus $\mu$ = $w_{1}$ and  1 - $\mu$ = $w_{2}$.\\
\end{itemize}

\subsection{Relation with the preferential attachment}

Since the centrality we propose takes inspiration from the preferential attachment phenomenon, we will now start by developing Eq.(\ref{equation4}) to establish that link. But first, let us agree on the following notations:\\

\begin{itemize}
\item $N_{i}$: The set of node i neighbors.
\item $N^{c}_{i}$: The complement of the set $N_{i}$.
\item V=\{$v_1$, $v_2$,..., $v_n$\}: The set of network nodes.
\item all = V - \{$v_{i}$\}: The set of all network nodes except the node i.
\item all= $N_{i}$  $\cup$ $N^{c}_{i}$ and $N_{i}$  $\cap$ $N^{c}_{i}$ = $\emptyset$
\item $N_{i}$ = all - $N^{c}_{i}$ 
\end{itemize}
By developing Eq.(\ref{equation4}):
\begin{gather*} 
(\sum_{j \in all} k_{j}) \Pi_{n\rightarrow i} (k_{i})= k_{i} \\
(\frac{\sum_{j \in all} k_{j}}{\sum_{j \in all} k_{j} - \sum_{j \in N^{c}_{i}} k_{j}}) \Pi_{n\rightarrow i} (k_{i})= \frac{k_{i}}{\sum_{j \in all} k_{j} - \sum_{j \in N^{c}_{i}} k_{j}} \\
(\frac{\sum_{j \in all} k_{j}}{\sum_{j \in N_{i}} k_{j}}) \Pi_{n\rightarrow i} (k_{i})= \frac{k_{i}}{\sum_{j \in N_{i}} k_{j}} 
\end{gather*} \\
We note $\frac{k_{i}}{\sum_{j \in N_{i}} k_{j}}$ by $\Pi_{n\rightarrow i}^{local} (k_{i})$ to obtain:
\begin{equation}
\Pi_{n\rightarrow i}^{local} (k_{i})=(\frac{\sum_{j \in all} k_{j}}{\sum_{j \in N_{i}} k_{j}}) \Pi_{n\rightarrow i} (k_{i}).
\label{eq:equation5}
\end{equation}
 Eq.(\ref{eq:equation5}) expresses the preferential attachment at a local level.  This result allows us to construct the local measure of M-Centrality.\\
By developing Eq.(\ref{eq:equation6}) we obtain:
\begin{gather*} 
\mathfrak{d_{i,j}}=  | N_{i}|.\Big|\frac{k_{j}}{\sum_{j \in N_{i}}k_{j}} - \frac{k_{i}}{\sum_{j \in N_{i}}k_{j}}\Big|\\
\mathfrak{d_{i,j}}=   | N_{i}|.\Big|\frac{k_{j}}{\sum_{j \in N_{i}}k_{j}} - (\frac{\sum_{j \in all} k_{j}}{\sum_{j \in N_{i}} k_{j}}) \Pi_{n\rightarrow i} (k_{i})\Big|
\end{gather*} \\
Replacing $(\frac{\sum_{j \in all} k_{j}}{\sum_{j \in N_{i}} k_{j}}) \Pi_{n\rightarrow i} (k_{i})$ by $\Pi_{n\rightarrow i}^{local} (k_{i})$ and $\frac{k_{j}}{\sum_{j \in N_{i}}k_{j}}$ by $\bar{k}_{j}^{N_{i}}$ gives:
\begin{equation}
\mathfrak{d_{i,j}} =   | N_{i}|.|\bar{k}_{j}^{N_{i}} - \Pi_{n\rightarrow i}^{local} (k_{i})|.
\label{eq:equation7}
\end{equation}
Finally, replacing $\Delta\mathfrak{D_{i}}$ by its value in Eq.(\ref{eq:equation9}) we obtain:
\begin{equation}
 M_{i}= \mu K_{s_{i}} + (1 - \mu) \sum_{j}  | N_{i}|.|\bar{k}_{j}^{N_{i}} - \Pi_{n\rightarrow i}^{local} (k_{i})|.
\label{equation10}
\end{equation}
This expression shows that the M-Centrality is linked to the preferential attachment phenomenon. 
\subsection{Computational complexity}\label{regsol0}

\begin{itemize}
\item The calculation of $K_s$ index for all nodes has a complexity of $\mathcal{O}(n)$, where n is the size of the graph.
\item The calculation of $\Delta\mathfrak{D}$ for all nodes has a complexity of $\mathcal{O}(n)$.
\item The calculation of the entropy matrix and the weight $\mu$  has a complexity of $\mathcal{O}(2n)$.
\item The calculation of M-Centrality for all nodes has a complexity of $\mathcal{O}(n)$ + $\mathcal{O}(n)$ + $\mathcal{O}(2n)$= $\mathcal{O}(n)$.
\end{itemize}

\section{Experimental results}
To evaluate the efficiency of the proposed method, eight real-world networks\footnote{The networks used can be found via this link: https://icon.colorado.edu/ and http://vlado.fmf.uni-lj.si/pub/networks/doc/erdos/ for Paul Erdős collaborations network} (small \cite{lusseau2003bottlenose, knuth1993stanford, beveridge2016network,batagelj2000some} and large scale \cite{adamic2005political,guimera2003self, newman2006finding, opsahl2011anchorage}) are studied. All datasets are considered undirected and unweighted; also, the largest connected component was used in the spreading process using the SIR model. The statistical properties relative to these networks are listed in Table \ref{table:1}. Concerning the evaluation process, it comprises the ability to detect key nodes\footnote{For all the Tables that follow, the nodes marked in color (blue, red, yellow, green, violet, orange and olive) are the most central according to literature. For networks with unknown information about central nodes, we mark in gray are the nodes that appear frequently in the ranking lists}, monotonicity, rank correlation, impact of nodes removal on network efficiency and finally the spreading capabilities of nodes. The results of the proposed centrality are compared to Gravity (Gr), DIL, Personalized PageRank (PPR)\footnote{The parameters used are: (1) a damping factor $\alpha = 0.15$ as experiments suggest that small changes in $\alpha$ have little effect in practice\cite{jeh2003scaling, srivastava2017discussion}. (2) for the preference vector, we choose to favor the Hubs.}, ClusterRank (CR) and Collective influence (COI) centralities.

\begin{table}[h!]
\renewcommand{\arraystretch}{1}
\caption{The statistical properties of the networks under study, where $<k>$ is the average degree of the network, $k_{max}$ the highest degree, $K_{s_{max}}$ the highest coreness, $\sigma$ the size of the giant component and $\beta_{th}$ the epidemic threshold.}
\label{table:1}
\centering
\resizebox{6.2in}{!}{
\begin{tabular}{|c |c  c  c c c c c c|}
    \hline
    Network & Type& Number of nodes & Number of edges& $<k>$ & $k_{max}$ & $K_{s_{max}}$ &$\beta_{th}$& $\sigma$\\
    \hline\hline
    Dolphins         & Social     &62 & 159& 5.12&  12  & 4  &0.139& 62                \\
    \hline
    Les Misérables  & \multirow{2}{*}{Co-appearance}       &77 & 254& 6.59& 36  &  9 &0.083&77            \\
    Game Of Thrones  &     &107 &352 & 6.57& 36  & 7  & 0.075&107           \\
    \hline
    Paul Erdős collaborations &\multirow{2}{*}{Collaboration}  &492 &1417& 5.76&  42  & 9 &0.058&  446               \\
    Netscience &  &1589 & 2742& 3.45& 34  & 19  &0.052 & 379                     \\
    \hline
    US Political blogs & Web-graph &1490 & 16715& 22.43& 351  & 36  &0.013&1222                     \\
    \hline
    E-mail     &   Communication    &1133& 5451& 9.62& 71 &  11 &0.048 & 1133                       \\ 
    \hline
    US airport 2010  & Traffic &1574 & 17215& 21.87& 314  &  64 &0.008 & 1572                 \\
   \hline
\end{tabular}
}
\end{table}

\subsection{Evaluation of nodes ranking} \label{sec:1}
\subsubsection{Small scale networks} 

Since there is no consensus on the concept of centrality, real networks with known information about the node's importance are commonly used as benchmarks to assess the effectiveness of centrality measures. We present the results of the experimental evaluation on three well documented networks (Dolphins, Les Misérables and Game Of Thrones) and a network with no information about central nodes (Paul Erdős collaborations). We choose to concentrate on these small scale networks in order to better understand the behavior of the centrality measures.

First, let's study the influence of the weight variation $\mu$ on the M-Centrality measure for the four networks. Tables.\ref{apptable1} and \ref{apptable2} illustrate the M-Centrality evolution of the nodes ranking for various values of the weight. We can clearly distinguish three major behaviors corresponding to the following cases $\mu$ = 0, 0 $<$ $\mu$ $<$1, and $\mu$ = 1. At the extremes, M-Centrality reduces to $\Delta\mathfrak{D}$ and Coreness while in between these two it tends to adapt to both local and global topological properties of the network. The major shortcoming of this is the fact that $\mu$ can vary in a wide range without impacting the nodes rankings. This is a very interesting result because it shows that a node needs to be strategically positioned in the network along with a high degree variation in its neighborhood in order to be considered influential. For more details about ranking, we suggest to refer to \ref{regsol}.

The next step is to determine the optimal weight value of the M-Centrality measure using the entropy weighted technique. The values obtained for the four networks are reported in \textit{Further analysis} section, Table \ref{table:11}. For Dolphins (resp. Les Misérables and Game of Thrones) the weight $\mu$ = 0.44 (resp. $\mu$ = 0.33 and $\mu$ = 0.36) puts more emphasis on the local measure $\Delta\mathfrak{D}$. The main reason for this is that Coreness assigns the same rank to many nodes. Therefore, its entropy is smaller than the one associated to the local measure $\Delta\mathfrak{D}$, and so is its weight in the M-centrality. This result makes sense. Indeed, $\Delta\mathfrak{D}$ is more efficient at identifying key characters of the novel. However, combining Coreness with degree variation allows to distinguish the characters that are grouped in the same category by the Coreness centrality. For Paul Erdős collaborations,  both global and local information are given the same importance by the entropy weighted technique.

After we study the behavior of M-Centrality based on $\mu$ values, we focus now on its capability of detecting key nodes and then compare the results with the alternative measures presented previously. For Dolphins, the information provided in \cite{lusseau2003bottlenose} reports that the female denoted SN100 plays an important role holding the community together. Table \ref{table:2} shows that DIL is the only centrality measure that puts SN100 at the top of the list. It is understandable given the fact that its disappearance split the network into two groups. ClusterRank, Personalized PageRank and Collective Influence are the only ones that fail in identifying SN100 in the top 15. In fact the first one ranks it at the $46^{th}$ position while the second and third ones rank it at the $19^{th}$ position. For M-Centrality, it is in the top 15. However, this result suggests that the measures do not exploit adequately the information about bridges that is well encoded in the DIL. Further improvement for M-Centrality needs to be made in this direction. 

For Les Misérables, The ex-convict, Jean Valjean, is a central character of the novel. He spends a great deal of time running away from Inspector Javert. He is also closely tied to his adopted daughter, Cosette, and her future husband, Marius. Fantine, Cosette's mother, Mr and Mme Thenardier and their son Gavroche are also important characters of the novel. We report on Table \ref{table:2} the top 15 nodes sorted by relevance. M-Centrality is the only measure that succeeds in ranking Valjean as the most influential character of the novel. Additionally, it identifies all the key characters of the novel in the top 15. For Gravity, DIL and Personalized PageRank, they all rank Gavroche, Valjean and Marius in the top 4. Thenardier and Javert are always in the top 15 but with very various rankings. We may also notice that Collective Influence and ClusterRank give the worst ranking as they clearly fail to detect main characters. The major shortcoming of these rankings is that (except for M-Centrality) Cosette does not appear in the top 15 major nodes and that the central hub (Valjean) is not necessarily the main bridge. 

\begin{table}[h!]
\renewcommand{\arraystretch}{1}
\caption{First 15 nodes sorted by relevance according to the centrality measures in the networks Dolphins and Les Misérables. M ($\mu = 0.44$) and M ($\mu = 0.33$) refer to the M-Centrality ranking corresponding to the weight obtained by the entropy weighted technique.}
\label{table:2}
\centering
\resizebox{6in}{!}{
\begin{tabular}{|c|| c| c |c |c|c |c|| c |c |c |c |c |c|}
   \hline 
    Networks          & \multicolumn{6}{c||}{Les Misérables}                                  & \multicolumn{6}{|c|}{Dolphins}  \\
   \hline\hline 
    Rank & M ($\mu = 0.44$) &Gr & DIL & CR&PPR&COI      & M ($\mu = 0.33$) &Gr & DIL & CR&PPR&COI   \\
    \hline\hline   
   1 & \cellcolor{blue}Valjean  &\cellcolor{red}Gavroche& \cellcolor{red}Gavroche& Bossuet& \cellcolor{red}Gavroche&Myriel&
         Grin&Grin&\cellcolor{blue}SN100&Hook&Grin&Beescratch                  \\
  
   2 & \cellcolor{red}Gavroche  & \cellcolor{blue}Valjean & Enjolras& Combeferre& \cellcolor{blue}Valjean&Listolier&       
        Trigger&SN4&Grin&MN105&SN4&PL\\
  
   3 & Myriel   & \cellcolor{pink}Marius & \cellcolor{pink}Marius& Feuilly& Enjolras&Fameuil&
        Topless&Topless&Topless&Jonah&Topless&SN4               \\
        
    4 & \cellcolor{pink}Marius & \ Enjolras& \cellcolor{blue}Valjean& Bahorel& \cellcolor{pink}Marius&Blacheville&
     Jet&Kringel&Web&MN83&Scabs&Trigger \\
     
     5 & \cellcolor{yellow}Javert &  Bossuet& Bossuet& Joly& Bossuet&Favourite&
     Web&SN9&Gallatin&Gallatin&Trigger&TR77  \\
     
     6 & \cellcolor{violet}Fantine & Courfeyrac& Courfeyrac& Enjolras& Courfeyrac&Dahlia&
  SN4&Scabs&Scabs&Topless&Patchback&Upbang         \\
  
   7 & \cellcolor{green}Thenardier  & \cellcolor{green}Thenardier& Bahorel& Courfeyrac& Bahorel  &Zephine&
  Scabs&\cellcolor{blue}SN100&SN4&Scabs&TR99&Oscar      \\
  
     8 & Enjolras & \cellcolor{yellow}Javert & Joly& Prouvaire& Joly&Bahorel&
      Patchback&TR99&Feather&Feather&SN9 &Scabs             \\
      
     9 & \cellcolor{olive}Cosette   &  Bahorel& Combeferre& Gueulmer& Combeferre&Joly&
  Kringel&Patchback&Trigger&SN90&Hook &DN63              \\
  
     10 &Bossuet & Joly& Feuilly& Babet&Feuilly &Fantine&
      SN63&Beescratch&MN105&DN21&MN105 &SN96            \\
      
     11 & \cellcolor{orange}Mme Thenardier  & Combeferre& Grantaire&Grantaire&Mabeuf &Courfeyrac&
      Beescratch&Jonah&Jonah&Grin&Jonah&SN63       \\
      
     12 &Montparnasse    & Feuilly&Mabeuf&Montparnasse&\cellcolor{green}Thenardier &Combeferre&
      Stripes&Trigger&Patchback&SN4&SN63&SN9               \\
      
     13 &Gueulemer  & Mabeuf& \cellcolor{green}Thenardier& Claquesous&\cellcolor{yellow}Javert &Feuilly&
     \cellcolor{blue}SN100&Double&DN21&Web&Kringel&Grin  \\   
     
     14 &Babet  & Grantaire&Prouvaire&Mabeuf&Grantaire&Mabeuf&
     Gallatin&Break&MN83&Upbang&MN83&Zap                 \\
     
     15 &Courfeyrac  & Gueulemer&\cellcolor{yellow}Javert&MmeHucheloup&Prouvaire&Grantaire&
    SN9&SN63&Upbang&SN9&Stripes&Topless      \\
   \hline
\end{tabular}
}
\end{table}

Moving on to the third network, Game Of Thrones, no one can deny the important role played by Tyrion Lannister in the story. Indeed, all the centralities (except ClusterRank) identify this character as the most important one. We also can see that M-Centrality performs the best as it identifies, along with Tyrion, a majority of other key characters, including his brother and sister Jamie and Cersei Lannister, Jon Snow, Catelyn stark and her three children Robb, Sansa and Arya and the mother of dragons Daenerys Targaryen. Collective influence manages to identify some major characters, with Daenerys ranked at the top of the list. However, we may notice the absence of Tyrion in the top 15. Again, ClusterRank gives the poorest ranking results. Overall, two important observations need to be made. First, the fact that only the proposed method succeeds in ranking Daenerys in the top 15 ($3^{rd}$ position). Second, for Jon Snow it is only M-Centrality that managed to give him importance by ranking him $2^{nd}$ most influential character.

Paul Erdős collaborations network is a collaboration graph of mathematicians where two mathematicians are joined by an edge whenever they co-authored a paper together. As the concept of  ''central'' nodes is relative, one may consider influence in term of the number of co-authors (degree), others may consider nodes located in the core, or those without the network will split in two or more sub-graphs, etc... This can intuitively seem a good approach, but in reality it tends to be biased since nodes that are considered central in term of connections (Hubs) in the topological context, can be less influential when talking about spreading capabilities where nodes located in the core are more important. 

Being said, at this stage of the paper and especially for networks with unknown information about central nodes, we make the assumption that all the rankings provided by the centralities under study are correct. Later, these rankings will be exploited in studying the impact of deleting keys nodes on network structure and efficiency and thus we can conclude which measure(s) provide(s) the most suitable ranking. From Table \ref{table:3}, the only observation we can make is that despite all centralities have different conceptions of influence, some nodes (marked in gray) appear frequently in all ranking results. There are also some concordances between the different centralities on ranking certain nodes in the top 15. For example M-Centrality and Gravity (resp. DIL, ClusterRank, Personnalized PageRank and Collective Influence) agree on 11 (resp. 7, 3, 8 and 1) out of 15 rankings. 

\begin{table}[h!]
\renewcommand{\arraystretch}{1}
\caption{First 15 nodes sorted by relevance according to the centrality measures in the networks Game Of Thrones and  Paul Erdős collaborations network. M ($\mu = 0.36$) and M ($\mu = 0.50$) refer to the M-Centrality ranking corresponding to the weight obtained by the entropy weighted technique.}
\label{table:3}
\centering
\resizebox{6in}{!}{
\begin{tabular}{|c|| c| c |c |c |c |c|| c |c |c |c |c |c|}
   \hline 
    Networks          & \multicolumn{6}{c||}{Game Of thrones}                                  & \multicolumn{6}{|c|}{Paul Erdős collaborations network}  \\
   \hline\hline 
    Rank & M ($\mu = 0.36$) &Gr & DIL & CR&PPR&COI     & M ($\mu = 0.50$) &Gr & DIL & CR&PPR&COI     \\
  \hline\hline
  1    &\cellcolor{blue}Tyrion&\cellcolor{blue}  Tyrion  &\cellcolor{blue}Tyrion & Eddard  &\cellcolor{blue}Tyrion&\cellcolor{pink}Daenerys
  &      HARARY, FRANK    &    \cellcolor{gray}GRAHAM, RONALD L.     &       \cellcolor{gray} RODL, VOJTECH     &SZEMEREDI, ENDRE        &\cellcolor{gray}GRAHAM, RONALD L.&POMERANCE, CARL\\
  
2     &  \cellcolor{red}Jon   &\cellcolor{yellow}Sansa   &\cellcolor{yellow}Sansa   &Meryn  &\cellcolor{yellow} Sansa&Mance
& \cellcolor{gray} GRAHAM, RONALD L. &            \cellcolor{gray}RODL, VOJTECH     &  \cellcolor{gray} GRAHAM, RONALD L.  & TROTTER, WILLIAM T., JR.     &       \cellcolor{gray} RODL, VOJTECH&HAJNAL, ANDRAS\\

3 & \cellcolor{pink}Daenerys   & Robb &  \cellcolor{violet}Jaime&    Ilyn &  \cellcolor{violet}Jaime&Samwell
&   POMERANCE, CARL &               \cellcolor{gray}ALON, NOGA        &        \cellcolor{gray}ALON, NOGA          & LEHEL, JENO             &   \cellcolor{gray}ALON, NOGA&CHARTRAND, GARY\\

4  &   \cellcolor{yellow}Sansa   &\cellcolor{violet}Jaime   & Robb &Joffrey  &  Robb&\cellcolor{green}Cersei
&       TUZA, ZSOLT       &       TUZA, ZSOLT            &  TUZA, ZSOLT                  &FRANKL, PETER      &   SPENCER, JOEL H.&HARARY, FRANK\\

5  &    Robb   &Tywin  &\cellcolor{green}Cersei&   Balon&  \cellcolor{green}Cersei&\cellcolor{blue}Tyrion
&       \cellcolor{gray}RODL, VOJTECH  &       SPENCER, JOEL H.     &    SPENCER, JOEL H.        &   JACOBSON, MICHAEL S.         &     TUZA, ZSOLT&STRAUS, ERNST G.\\

6  &   Tywin   & \cellcolor{orange}Arya   & \cellcolor{orange}Arya   &Petyr   &Tywin& \cellcolor{red}Jon
 &     SOS, VERA T.      &      HARARY, FRANK        &   FUREDI, ZOLTAN            &   \cellcolor{gray}RODL, VOJTECH      &     FUREDI, ZOLTAN&TUZA, ZSOLT\\

7  &   \cellcolor{violet}Jaime  &\cellcolor{green}Cersei &Joffrey & Gregor &Joffrey&Jorah
   &     \cellcolor{gray} ALON, NOGA     &      FUREDI, ZOLTAN       &CHUNG, FAN RONG K.        & GOULD, RONALD J. &      CHUNG, FAN RONG K.&GYARFAS, ANDRAS\\

8 &  Samwell  &Robert &  Tywin &\cellcolor{green} Cersei &   \cellcolor{orange}Arya&Sandor
&  SPENCER, JOEL H.  &     CHUNG, FAN RONG K.     &     GYARFAS, ANDRAS  &  FUREDI, ZOLTAN   &      SZEMEREDI, ENDRE&  \cellcolor{gray}RODL, VOJTECH\\

9 &  Catelyn &Joffrey&  Robert  & Aerys & Robert&Rhaegar
 &   HAJNAL, ANDRAS   &        BOLLOBAS, BELA       &  SZEMEREDI, ENDRE       &SAKS, MICHAEL E.  &         LOVASZ, LASZLO&SOS, VERA T.\\

10 &  \cellcolor{green}Cersei &Catelyn& Catelyn  &Sandor& Catelyn&Joffrey
&   BOLLOBAS, BELA   &     FAUDREE, RALPH J.     &   FAUDREE, RALPH J.      &   \cellcolor{gray}ALON, NOGA        &      PACH, JANOS&BABAI, LASZLO\\

11  &  Mance &Stannis  &Sandor   & Arya  &Sandor&Beric
&       PACH, JANOS      &       SOS, VERA T.             & PACH, JANOS                 &  \cellcolor{gray}GRAHAM, RONALD L.   &        HARARY, FRANK&TURAN, PAL\\

12 &    \cellcolor{orange}Arya & Eddard  &Eddard   &Jaime& Stannis&Davos
&  STRAUS, ERNST G.     &         PACH, JANOS         &  LOVASZ, LASZLO         &  KUBICKA, EWA MARIE   &         BABAI, LASZLO&SPENCER, JOEL H.\\

13  & Robert  & \cellcolor{red}  Jon    & \cellcolor{red}Jon& Pycelle&  Eddard&\cellcolor{violet}Jaime
&  KLEITMAN, DANIEL J.    &     SZEMEREDI, ENDRE         &   BABAI, LASZLO    &  KUBICKI, GRZEGORZ    &      GYARFAS, ANDRAS&SARKOZY, ANDRAS\\

14 & Joffrey&  Sandor  &Gregor   &\cellcolor{yellow}Sansa  &Gregor&Tywin
& CHARTRAND, GARY     &      LOVASZ, LASZLO         &   FRANKL, PETER     & GYARFAS, ANDRAS     &   FAUDREE, RALPH J.&LOVASZ, LASZLO\\

15  &   Bran  &Gregor& Stannis& Stannis  &   \cellcolor{red}Jon&Edmure
& CHUNG, FAN RONG K.   &        HAJNAL, ANDRAS    & JACOBSON, MICHAEL S.  &   SCHELP, RICHARD H.    &        FRANKL, PETER&RENYI, ALFRED A.\\
   \hline
\end{tabular}
}
\end{table}

Now we turn to the comparative evaluation of the centrality measures. Table \ref{table:4} shows the monotonicity of the ranking methods. Results show that M-Centrality, Gravity and Personalized PageRank clearly outperform the other measures. In other words, few nodes are assigned the same rank. DIL  is more competitive than ClusterRank centrality except  in the case of Game Of Thrones network. Collective Influence also gives good performances in distinguishing between nodes importance.

Table \ref{table:5} reports the rank correlation of the various centrality measures. The main result is that M-Centrality is highly and positively correlated with Gravity, with $\tau$ values ranging from 0.71 to 0.85. This is reasonable since they are both a variant of Coreness centrality. Globally, the proposed method is moderately correlated with ClusterRank. The lowest correlation value is registered between M-Centrality and Collective Influence ($\tau_{M,COI}$ = 0.46) in the case of Les Misérables network. In other words, Collective Influence centrality behaves very differently than the proposed method for this network. Note that it is evident if we refer to the top fifteen nodes. Further investigations about the nature of the relation between the ranking produced by the M-Centrality and the one of its alternatives are reported in \ref{regsol2}.

\begin{table}[h!]
\renewcommand{\arraystretch}{1}
\caption{Monotonicity $\mathfrak{M}$ of centrality measures for the small scale real-world networks. The weight of the M-Centrality  is computed by the entropy weighted technique.}
\label{table:4}
\centering
\resizebox{4in}{!}{
\begin{tabular}{|c| c  c  c c c c|}
    \hline
    Network & $\mathfrak{M}$(M) & $\mathfrak{M}$(Gr) & $\mathfrak{M}$(DIL) & $\mathfrak{M}$(CR) & $\mathfrak{M}$(PPR) & $\mathfrak{M}$(COI)  \\
    \hline\hline
   Dolphins & 0.989 &  0.997& 0.958&  0.873& 0.997&0.960\\
   Les Misérables &0.958  & 0.958 &  0.876& 0.854 &0.958&.864 \\
   Game of Thrones   & 0.993& 0.994& 0.937& 0.948&0.995&0.953\\
   Paul Erdős collaborations &0.982 & 0.987& 0.922&0.762 &0.987&0.879\\
 \hline
\end{tabular}
}
\end{table}

\begin{table}[h!]
\renewcommand{\arraystretch}{1}
\caption{Kendall’s tau ($\tau$) rank correlation coefficient for the small scale real-world networks.}
\label{table:5}
\centering
\resizebox{4in}{!}{
\begin{tabular}{|c| c  c  c c c|}
    \hline
    Network               & $\tau$(M, Gr) & $\tau$(M, DIL)& $\tau$(M, CR) & $\tau$(M, PPR)& $\tau$(M, COI)     \\
    \hline\hline
   Dolphins                &0.716 & 0.699& 0.589& 0.504&0.517\\
   Les Misérables        & 0.824& 0.703& 0.632& 0.736&0.467\\
   Game of Thrones   & 0.857&0.820 & 0.612& 0.748&0.566\\
   Paul Erdős collaborations &0.856 &0.623 & 0.648&0.744&0.857\\
  \hline
\end{tabular}
}
\end{table}

\subsubsection{Large scale networks}

After we study the behavior of M-Centrality on small scale networks, we move on to present the results of the experimental evaluation on four large scale networks. It includes E-mail, Netscience, US airport and US Political blogs. As the performances of the proposed measure have been established previously, it will be interesting to test its effectiveness on much larger graphs that are not necessarily well documented. Details concerning the impact of the weight variation on M-Centrality are reported in \ref{regsol}, Tables \ref{apptable3} and \ref{apptable4}. 

First, we determine the key nodes identified by the different centralities. Tables \ref{table:6} and \ref{table:7} present our results. The main observation is that each centrality identifies various key nodes. However, some nodes appear more frequently in the top 15 of multiple centralities, this can possibly suggest their potential importance. The first network, E-mail, represents the exchange of emails among members of the Rovira i Virgili University in Spain, in 2003. The nodes marked in gray appear in 4 out of 5 ranking results except ClusterRank. However, an important thing to notice about these nodes is the fact that none of them is strategically located in the network. In other words, they are not in the core of the network. 

Netscience is a network of co-authorships between scientists whose research centers on the properties of networks. Edges join every pair of individuals whose names appear together as authors of a paper. The nodes marked in gray are the ones who appear in all ranking results except M-Centrality and Collective Influence. In addition, they all belong to the core of the network. This result suggests that in this network, the proposed measure and Collective Influence quantify node influence differently from its alternatives. Note that these two methods agree both on ranking BARABASI, A, NEWMAN, M and JEONG,H in the top 3.

\begin{table}[h!]
\renewcommand{\arraystretch}{1}
\caption{First 15 nodes sorted by relevance according to the centrality measures in the networks E-mail and  Netscience.}
\label{table:6}
\centering
\resizebox{6in}{!}{

\begin{tabular}{|c|| c| c |c |c |c |c||c |c |c |c |c |c|}
   \hline 
    Networks          & \multicolumn{6}{c||}{E-mail}                                  & \multicolumn{6}{|c|}{Netscience}                      \\
   \hline\hline 
    Rank & M ($\mu = 0.40$) &Gr & DIL & CR&PPR&COI     & M ($\mu = 0.50$) &Gr & DIL & CR&PPR&COI      \\
   \hline\hline
     1          &\cellcolor{gray}105 &\cellcolor{gray}105 &\cellcolor{gray}105 &886 &\cellcolor{gray}105& 16&           
       BARABASI, A        &  UETZ, P         & UETZ, P        &  \cellcolor{gray}GIOT, L  &UETZ, P&NEWMAN, M\\

     2          &23 &333  &\cellcolor{gray}16 &888&  \cellcolor{gray}16& 105&           
      NEWMAN, M      &  CAGNEY, G      &  CAGNEY, G        &\cellcolor{gray}JUDSON, R   &CAGNEY, G&BARABASI, A\\

     3          & 333  &\cellcolor{gray}42 &299 &887 &\cellcolor{gray}196& 24&         
      JEONG, H   &  MANSFIELD, T  &   MANSFIELD, T   &     \cellcolor{gray}KNIGHT, J & MANSFIELD, T& JEONG, H\\

     4          &41  &23 &\cellcolor{gray}196& 788 &204&   564&          
              YOUNG, M        &  \cellcolor{gray}GIOT, L      &   \cellcolor{gray} GIOT, L    &  \cellcolor{gray}LOCKSHON, D &\cellcolor{gray}GIOT, L&SOLE, R\\

     5         & \cellcolor{gray}16&  76 &  3 &571  &\cellcolor{gray}42& 14&      
    BOCCALETTI, S    &    \cellcolor{gray}JUDSON, R       & \cellcolor{gray}JUDSON, R &      \cellcolor{gray}NARAYAN, V & \cellcolor{gray}JUDSON, R&HOLME, P  \\

     6         & \cellcolor{gray}42  &41  &\cellcolor{gray}42 &885  &49&  434&       
OLTVAI, Z        &\cellcolor{gray}KNIGHT, J       & \cellcolor{gray}KNIGHT, J   & \cellcolor{gray}SRINIVASAN, M  & \cellcolor{gray} KNIGHT, J&MORENO, Y\\

     7         & 233 &233 &204 &299  &56&  196&           
SOLE, R   &  \cellcolor{gray} LOCKSHON, D  &   \cellcolor{gray} LOCKSHON, D  &     \cellcolor{gray}POCHART, P &   \cellcolor{gray}LOCKSHON, D&OLTVAI, Z\\

     8         & 24  &52 &205& 426 &116& 72&      
ALON, U      & \cellcolor{gray}NARAYAN, V      & \cellcolor{gray}NARAYAN, V  &\cellcolor{gray}QURESHIEMILI, A   &\cellcolor{gray}NARAYAN, V&LATORA, V\\

     9         & 14   &3 &389  & 3 &333&   204&            
KURTHS, J   & \cellcolor{gray}SRINIVASAN, M    &\cellcolor{gray}SRINIVASAN, M        &   \cellcolor{gray} LI, Y  & \cellcolor{gray}SRINIVASAN, M&PASTORSATORRAS, R\\

    10        &\cellcolor{gray}196& \cellcolor{gray}196& 552  & 9  & 3&  354&         
DIAZGUILERA, A    &  \cellcolor{gray} POCHART, P     &  \cellcolor{gray}POCHART, P       &\cellcolor{gray} GODWIN, B   &\cellcolor{gray}POCHART, P&CALDARELLI, G\\
              
    11  &21 &135 &434 &205  &23&396&
     LATORA, V  &\cellcolor{gray}QURESHIEMILI, A&  \cellcolor{gray}QURESHIEMILI, A   &    \cellcolor{gray}CONOVER, D  &\cellcolor{gray}QURESHIEMILI, A&VAZQUEZ, A\\

    12 &355 &332 &332 &389 &128&116&
       HU, G           & \cellcolor{gray}LI, Y          &  \cellcolor{gray}LI, Y   &\cellcolor{gray}KALBFLEISCH, T & \cellcolor{gray}LI, Y&VESPIGNANI, A\\
    
    13 &578&  \cellcolor{gray}16 &726 &210 & 21&69&
   MUTH, S     &   \cellcolor{gray}GODWIN, B   &     \cellcolor{gray}GODWIN, B & VIJAYADAMODAR, G & \cellcolor{gray}GODWIN, B&DIAZGUILERA, A\\

    14 &135& 299&  56 &386 &206&21&
 VICSEK, T     &  \cellcolor{gray}CONOVER, D    &   \cellcolor{gray}CONOVER, D  &        YANG, M &   \cellcolor{gray}CONOVER, D& STAUFFER, D\\

    15  &76  &21  &49 &215  &41& 341&
      KAHNG, B   &\cellcolor{gray}KALBFLEISCH, T   &\cellcolor{gray}KALBFLEISCH, T     & JOHNSTON, M  &\cellcolor{gray}KALBFLEISCH, T&BOCCALETTI, S\\
   \hline
\end{tabular}
}
\end{table}

US airport is a complete network of flights among all commercial airports in the United States, in 2010, derived from the U.S. Bureau of Transportation Statistics (BTS) Transtats site. Weights represent the number of seats available on the flights between a pair of airports. In this network, all centralities (except ClusterRank) agree on ranking the nodes marked in gray in the top 15. Again, they are all located in the core of the network. 

The last network, US Political blogs, is a network of hyperlinks between web blogs on US politics, recorded in 2005 by Adamic and Glance. As for US airport and E-mail networks, all centralities (except ClusterRank) agree on ranking the nodes marked in gray in the top 15, with Gravity and DIL ranking them in the top 5. These nodes are located in the core of the network except for instapundit.com that has a coreness value of $K_s = 32$.

\begin{table}[h!]
\renewcommand{\arraystretch}{1.2}
\caption{First 15 nodes sorted by relevance according to the centrality measures in the networks US airport and Political blogs.}
\label{table:7}
\centering
\resizebox{6in}{!}{
\begin{tabular}{|c|| c| c |c| c |c |c|| c |c |c |c |c |c|}
   \hline 
    Networks          & \multicolumn{6}{c||}{US airport}                                & \multicolumn{6}{|c|}{US Political blogs}\\
   \hline\hline 
    Rank & M ($\mu = 0.67$) &Gr & DIL & CR&PPR&COI      & M ($\mu = 0.50$) &Gr & DIL & CR&PPR&COI      \\
   \hline\hline
1  & 766 &\cellcolor{gray}1200 &\cellcolor{gray}1200& 1201&\cellcolor{gray} 1200&152&
   blogsforbush.com                       &\cellcolor{gray}dailykos.com      &\cellcolor{gray}dailykos.com         &nielsenhayden.com/electrolite     &\cellcolor{gray}dailykos.com&blogsforbush.com\\
        
2   &\cellcolor{gray}114&  \cellcolor{gray}114 & \cellcolor{gray}114 &1391&  \cellcolor{gray}114&567&
\cellcolor{gray}dailykos.com     &           \cellcolor{gray}atrios.blogspot.com  &  \cellcolor{gray} atrios.blogspot.com   &                  michaelberube.com  &  \cellcolor{gray}atrios.blogspot.com&gevkaffeegal.typepad.com/\texttt{the\_alliance}\\

3  & 877 & \cellcolor{gray}435 & 709  &683 & 709&977&
drudgereport.com             & \cellcolor{gray}talkingpointsmemo.com   &\cellcolor{gray}talkingpointsmemo.com                  &  tbogg.blogspot.com       & \cellcolor{gray}talkingpointsmemo.com&liberaloasis.com\\

4  & 709 &\cellcolor{gray}1068 &\cellcolor{gray}1068  &759 & \cellcolor{gray}435&844&
\cellcolor{gray} instapundit.com            &        \cellcolor{gray}instapundit.com    & \cellcolor{gray} washingtonmonthly.com    &                 roadtosurfdom.com  &   \cellcolor{gray}washingtonmonthly.com&corrente.blogspot.com\\

5  &\cellcolor{gray}1200&  709&  \cellcolor{gray}435& 1011& \cellcolor{gray}1068&402&
\cellcolor{gray}    talkingpointsmemo.com         & \cellcolor{gray}    washingtonmonthly.com &   \cellcolor{gray}instapundit.com                &     aintnobaddude.com &     liberaloasis.com&pacificviews.org\\
 
6 & 1016&\cellcolor{gray}  391 & \cellcolor{gray}391  &239  &711&1660&
                    \cellcolor{gray}atrios.blogspot.com    &        digbysblog.blogspot.com  &  liberaloasis.com &                     busybusybusy.com   &    \cellcolor{gray}instapundit.com&bodyandsoul.typepad.com\\

7 &  \cellcolor{gray}500 &1252&  711& 1820 &\cellcolor{gray} 391&47&
                     powerlineblog.com                   &    juancole.com &    digbysblog.blogspot.com           &    bodyandsoul.typepad.com   & digbysblog.blogspot.com&busybusybusy.com\\

8   &389  &711 &1252  &169  &\cellcolor{gray}500&1186&
                     michellemalkin.com   &                    talkleft.com    &    bodyandsoul.typepad.com    &             atrios.blogspot.com/   &      bodyandsoul.typepad.com&seetheforest.blogspot.com\\

9 &  711 & 389 & 389 &  75 &1252&91&
                    truthlaidbear.com             &      liberaloasis.com            & talkleft.com        &     nomoremister.blogspot.com     &  pandagon.net&wampum.wabanaki.net\\
                    
10 &\cellcolor{gray}1068&  \cellcolor{gray}500 &\cellcolor{gray} 500  &760&  389&1677&
     \cellcolor{gray}           washingtonmonthly.com     &      madkane.com/notable.html    &      pandagon.net    &               wampum.wabanaki.net    &  talkleft.com&atrios.blogspot.com/ \\

11  &\cellcolor{gray}391  &982 & 206 &1320 & 877&726&
       littlegreenfootballs.com/weblog           &       powerlineblog.com  &     politicalstrategy.org              &       thetalkingdog.com  &    juancole.com&blogsagainsthillary.com\\

12  &215 & 206 & 982& 1628&  982&482&
                       wizbangblog.com     &                  pandagon.net     &        corrente.blogspot.com        &          nathannewman.org/log   &      politicalstrategy.org&coxandforkum.com\\

13  &\cellcolor{gray}435 & 311&  877& 1376 & 206&1190&
                        hughhewitt.com&       yglesias.typepad.com/matthew    &       tbogg.blogspot.com      &        elayneriggs.blogspot.com   &    corrente.blogspot.com&homespunbloggers.blogspot.com\\

14 & 875& 1353 & 311  &478  &215&1742&
                          juancole.com             &    tbogg.blogspot.com          &   prospect.org/weblog                       &    billmon.org      &  tbogg.blogspot.com&patriotboy.blogspot.com\\

15  &685 & 877  &215  &997  &311&836&
       lashawnbarber.com    &             michellemalkin.com  &     dneiwert.blogspot.com    &                      leanleft.com   &  dneiwert.blogspot.com&techievampire.net/wppol\\
   
   \hline
\end{tabular}
}
\end{table}

Next we examine the monotonicity of the different methods. The results are reported in Table \ref{table:8}. Again M-Centrality and Gravity clearly offer the best performances in all networks with a very high monotonicity score outperforming the one of the other centralities. We also notice the poor results of Personalized PageRank and Collective Influence in Netscience network, which can be interpreted as a sign of incapability to distinguish between nodes influence.

Table \ref{table:9} shows the correlation between M-Centrality and other centralities. One can see that in the four networks, M-Centrality is very correlated with Gravity. We also notice that the proposed measure exhibits poor relation with Personalized PageRank, especially in Netscience network ($\tau$(M, PPR) = 0.29). For the results concerning the concordance in high ranks between the ranking list produced by M-Centrality and the one of its alternatives, we suggest the reader to refer to \ref{regsol2}

\begin{table}[h!]
\renewcommand{\arraystretch}{1}
\caption{Monotonicity $\mathfrak{M}$ of the centrality measures for the large scale networks.}
\label{table:8}
\centering
\resizebox{4in}{!}{
\begin{tabular}{|c |c  c  c c c c|}
    \hline
    Network & $\mathfrak{M}$(M) & $\mathfrak{M}$(Gr) & $\mathfrak{M}$(DIL) & $\mathfrak{M}$(CR) & $\mathfrak{M}$(PPR)& $\mathfrak{M}$(COI)   \\
    \hline\hline
   Netscience         & 0.910 &  0.915& 0.803& 0.805& 0.016&0.275 \\
   US airport          &0.997 &  0.998&  0.897& 0.882& 0.998&0.915 \\
   E-mail                 & 0.998&  0.999& 0.961 &0.870 & 0.999&0.964 \\
   US Political blogs    & 0.936& 0.936 & 0.918 & 0.794& 0.937&0.859 \\
 \hline
\end{tabular}
}
\end{table}

\begin{table}[h!]
\renewcommand{\arraystretch}{1}
\caption{Correlation between M-Centrality and its alternatives for the large scale networks.}
\label{table:9}
\centering
\resizebox{4in}{!}{
\begin{tabular}{|c |c  c  c c c|}
    \hline
    Network               & $\tau$(M, Gr) & $\tau$(M, DIL)& $\tau$(M, CR) & $\tau$(M, PPR)& $\tau$(M, COI)   \\
    \hline\hline
   Netscience          & 0.883 & 0.841 & 0.806 & 0.297&0.500  \\
   US airport           & 0.867 &0.811  &0.711 &0.617&0.742  \\
   E-mail                  &0.881  & 0.527 & 0.569& 0.766&0.906 \\
   US Political blogs  &0.962  & 0.750 & 0.760& 0.885&0.756 \\
 \hline
\end{tabular}
}
\end{table}

\subsection{Impact of nodes removal on network efficiency}
A high monotonicity score alone does not mean necessarily the performance of a ranking method, that is why in many studies about ranking influential nodes, the ranking list of different centrality measures is evaluated in the context of network vulnerability and transmission dynamics. In this work we explore the two ways.

First, we evaluate the efficiency of the proposed method by examining the impact of removing the top most important nodes on network structure. Table \ref{table:10} shows the rest graph obtained after deleting the key nodes identified by each centrality. The main observation is that M-Centrality outperforms the other centralities, with the exception of the two collaborations networks where Collective Influence seems to be more effective. Indeed the removal of the most important nodes identified by the proposed measure has high impact on network structure.  These results suggest that the ranking list of the proposed measure is more consistent and accurate than the one of its alternatives.

\begin{table}[h!]
\renewcommand{\arraystretch}{1}
\caption{Number of connected components $\mathfrak{C}$ obtained after removing the most important nodes according to ranking lists produced by the various centrality measures.}
\label{table:10}
\centering
\resizebox{4in}{!}{
\begin{tabular}{|c| c|  c  c c c c c|}
    \hline
    Network & Number of removed nodes &$\mathfrak{C}$(M) & $\mathfrak{C}$(Gr) & $\mathfrak{C}$(DIL) & $\mathfrak{C}$(CR)& $\mathfrak{C}$(PPR)& $\mathfrak{C}$(COI)    \\
    \hline\hline
    Dolphins & 15&      12 & 11& 10&5&7&6 \\
    Les Misérables & 15&         26 & 15&14 &2&14&9 \\
    Game Of Thrones &20 &            29 & 22& 19& 5&22&23\\
    Paul Erdős collaborations &30 &           63 &58 & 58&48&61&70 \\
    Netscience & 50 &          462& 411&406 &395&418&490 \\
    US airport 2010 &50 &          286 & 203& 203&17&203&94 \\
    E-mail &50 &               42 & 21&17 &12&22&36 \\
    US Political blogs &50 &            362 & 327& 292& 274&296& 314\\
    \hline
\end{tabular}
}
\end{table}

\begin{figure*}[!h]
\centering
\subfigure{
\includegraphics[width=2.35in]{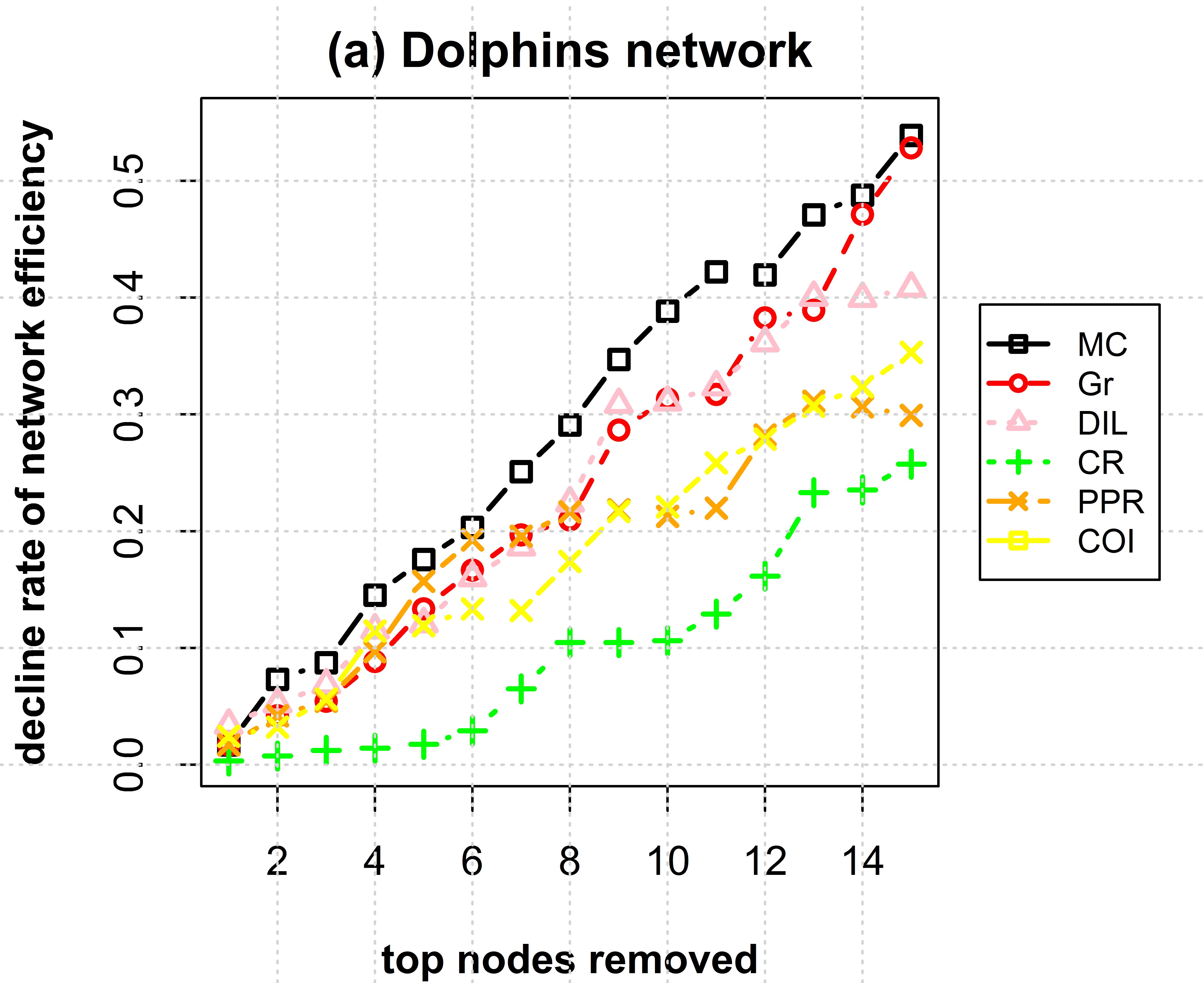}
\label{fig:1}}
\hfil
\subfigure{
\includegraphics[width=2.35in]{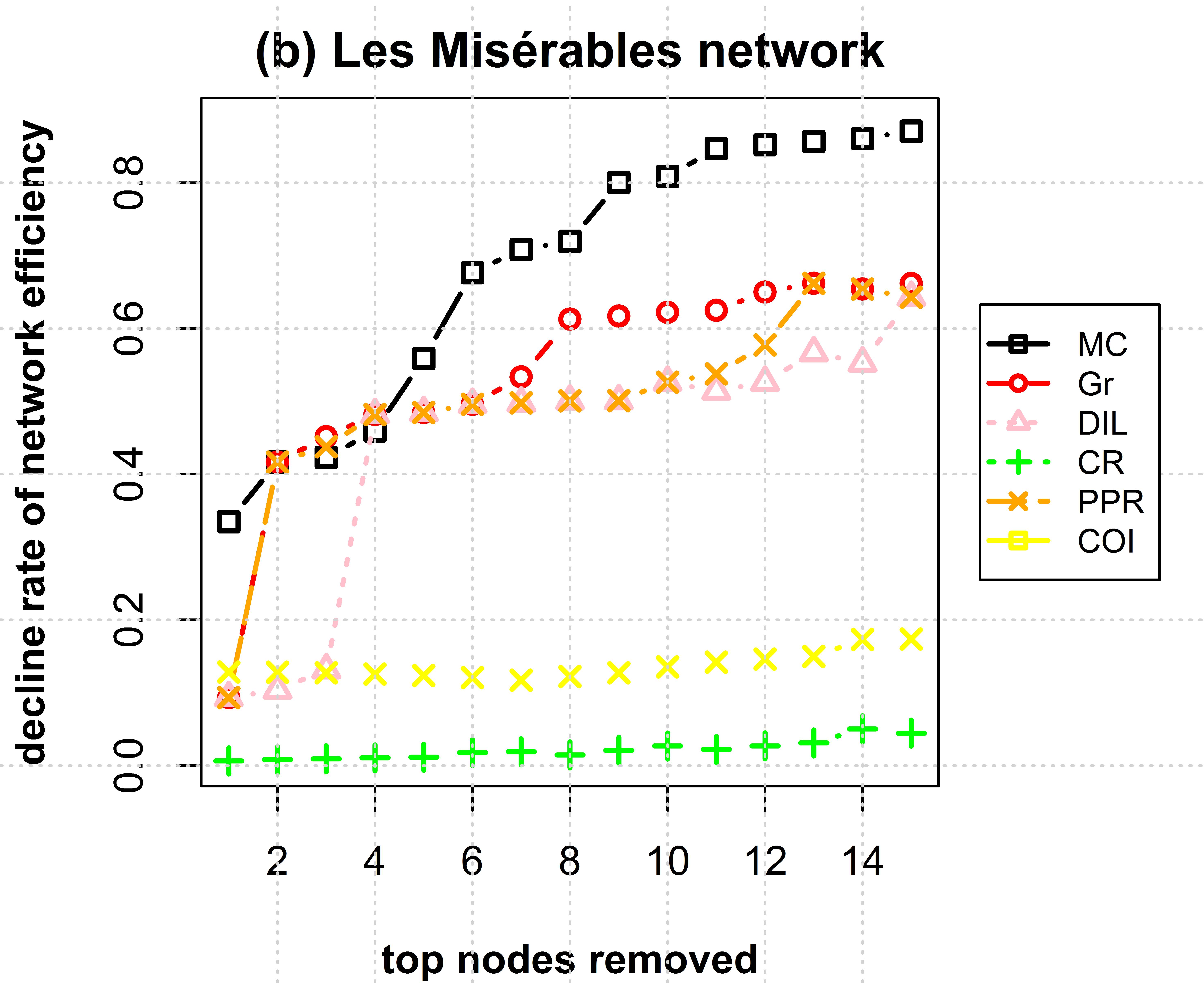}
\label{fig:2}}
\hfil
\subfigure{
\includegraphics[width=2.35in]{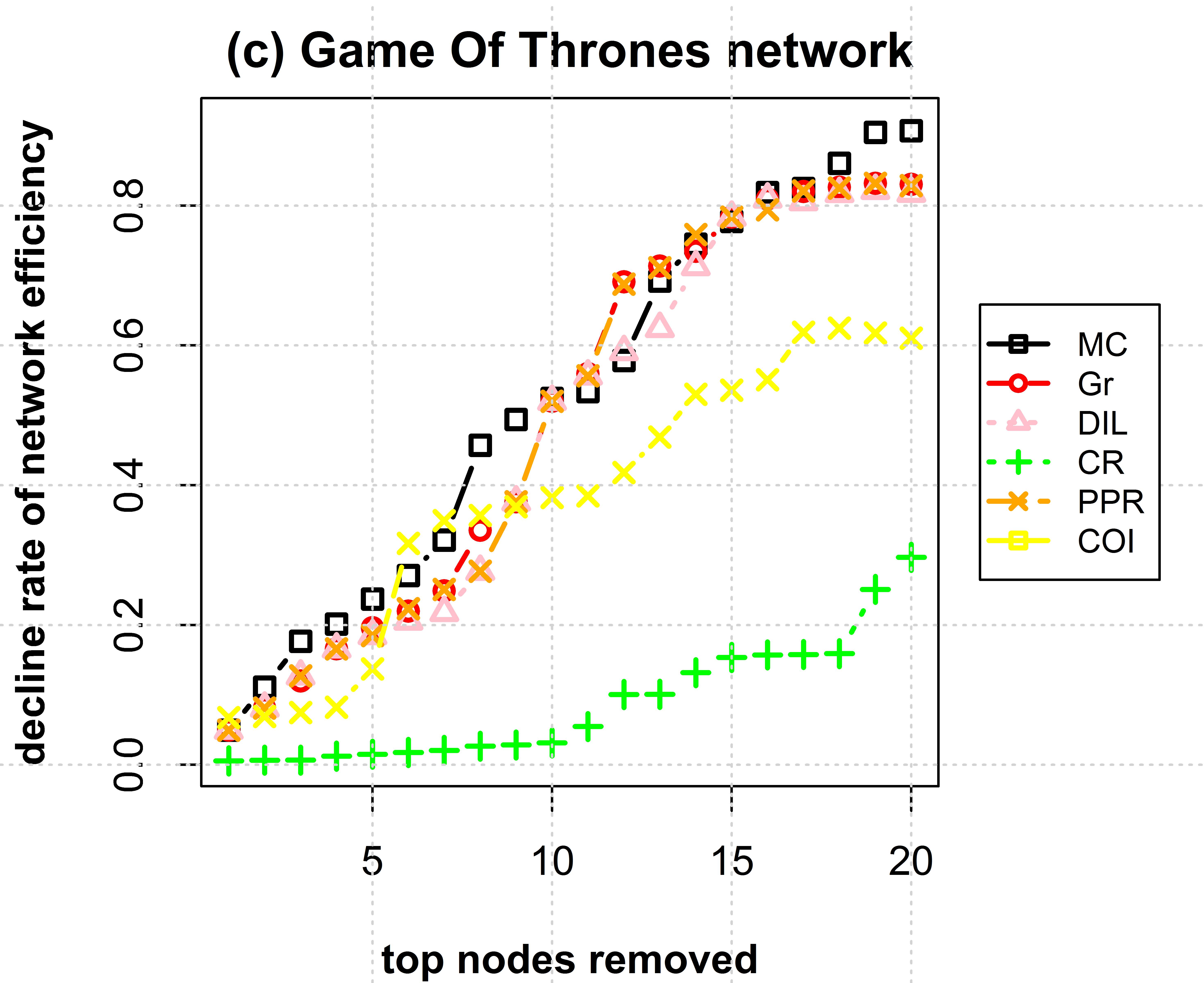}
\label{fig:3}}
\hfil
\subfigure{
\includegraphics[width=2.35in]{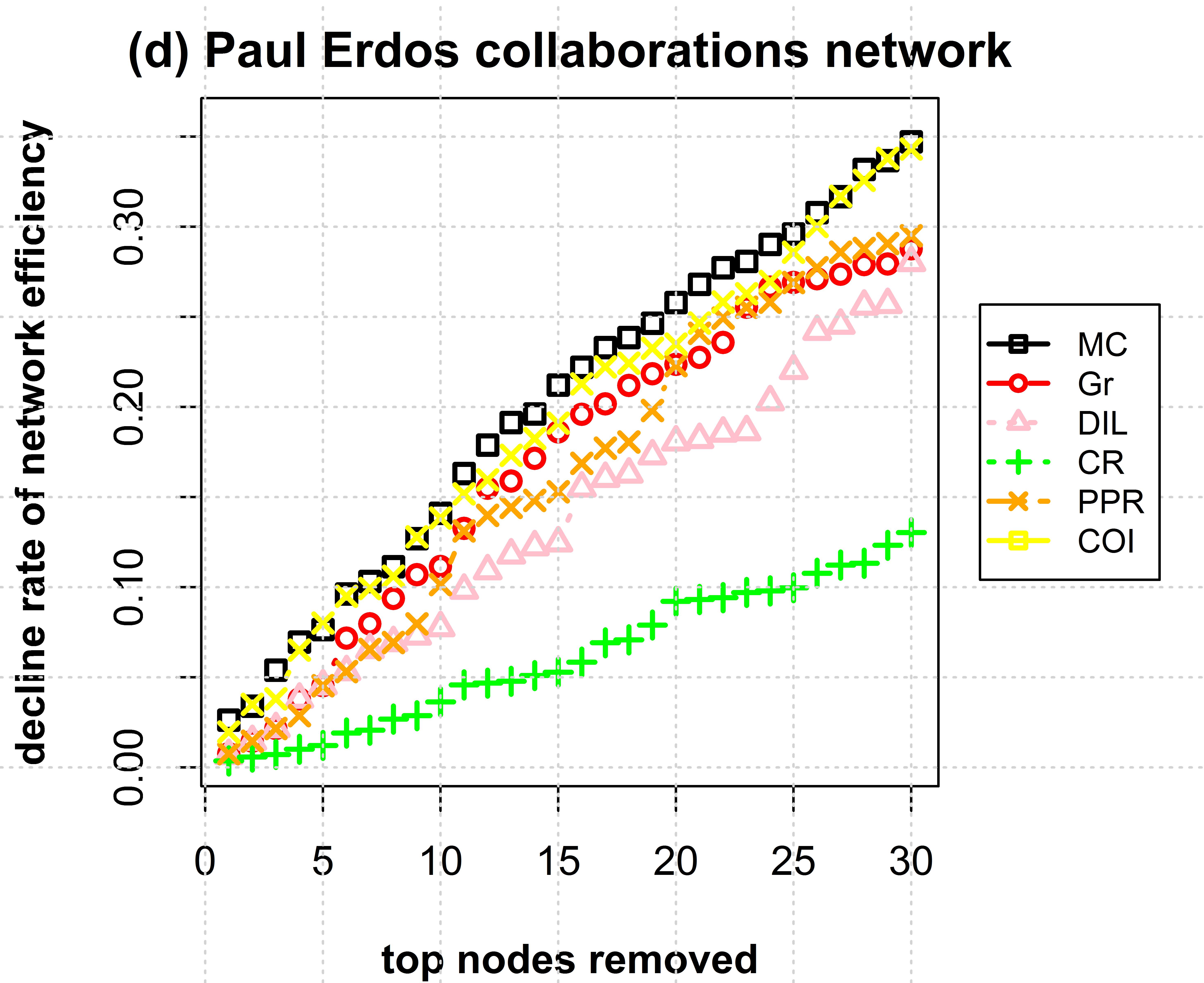}
\label{fig:4}}
\hfil
\subfigure{
\includegraphics[width=2.35in]{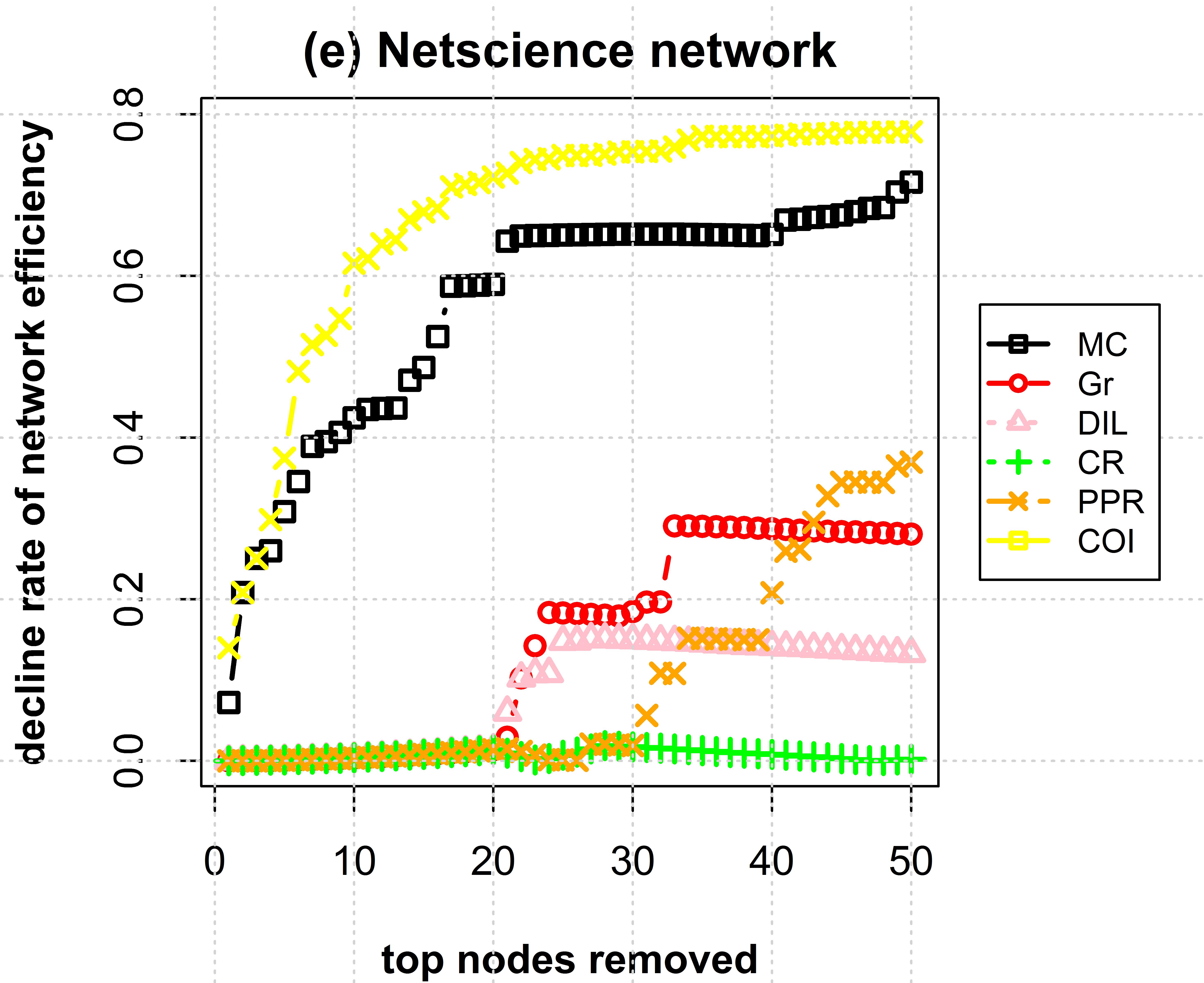}
\label{fig:5}}
\hfil
\subfigure{
\includegraphics[width=2.35in]{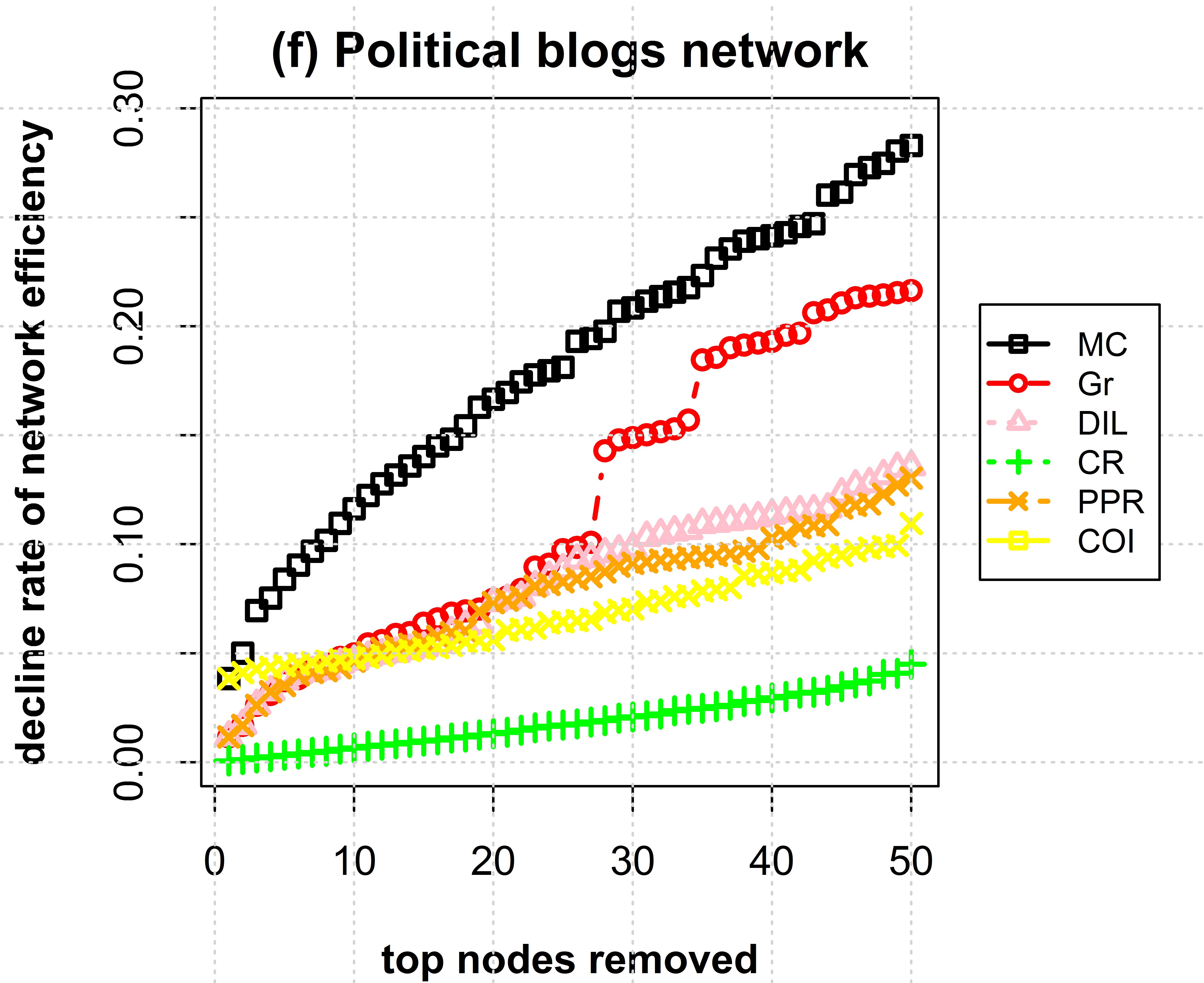}
\label{fig:6}}
\hfil
\subfigure{
\includegraphics[width=2.35in]{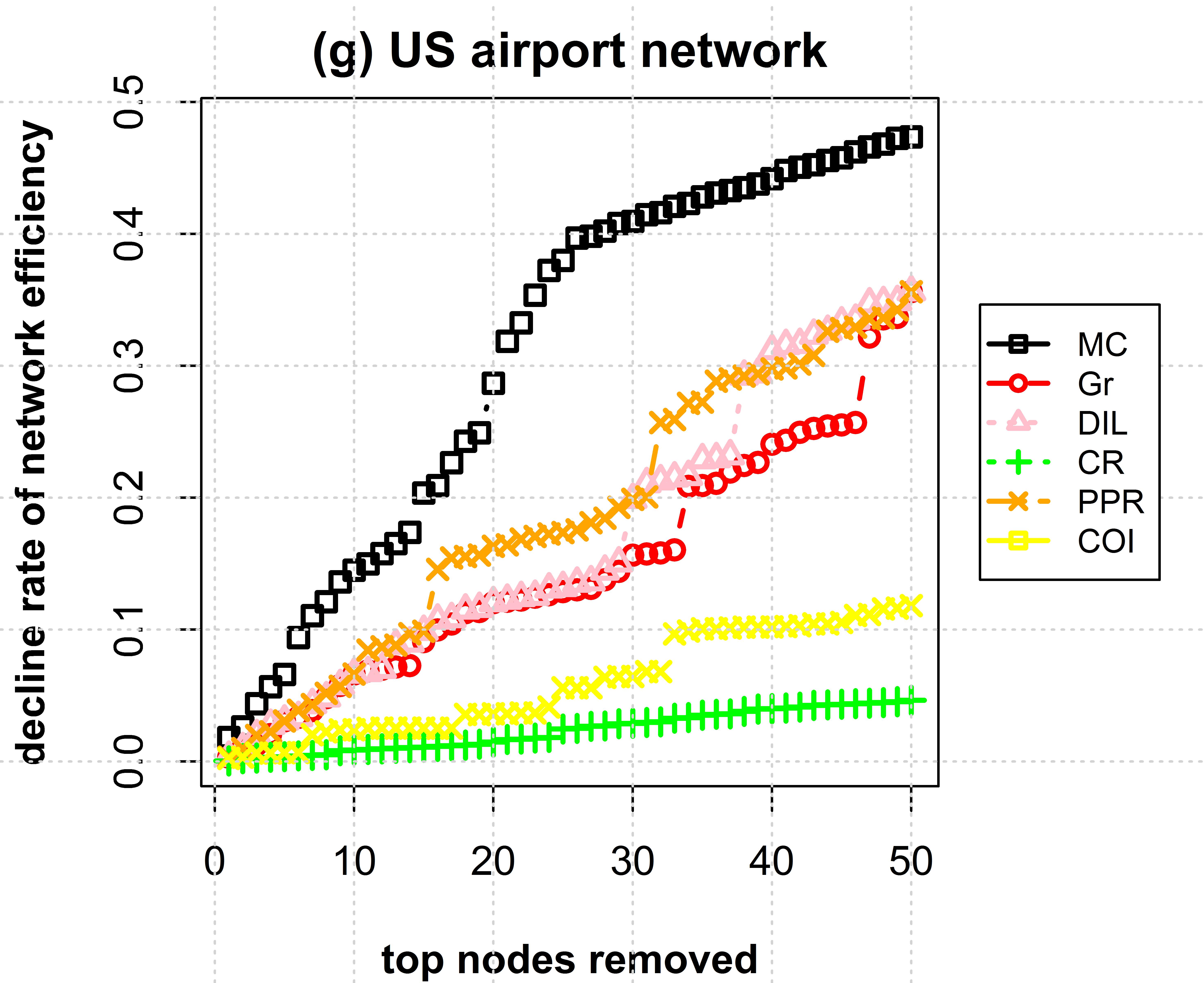}
\label{fig:7}}
\hfil
\subfigure{
\includegraphics[width=2.35in]{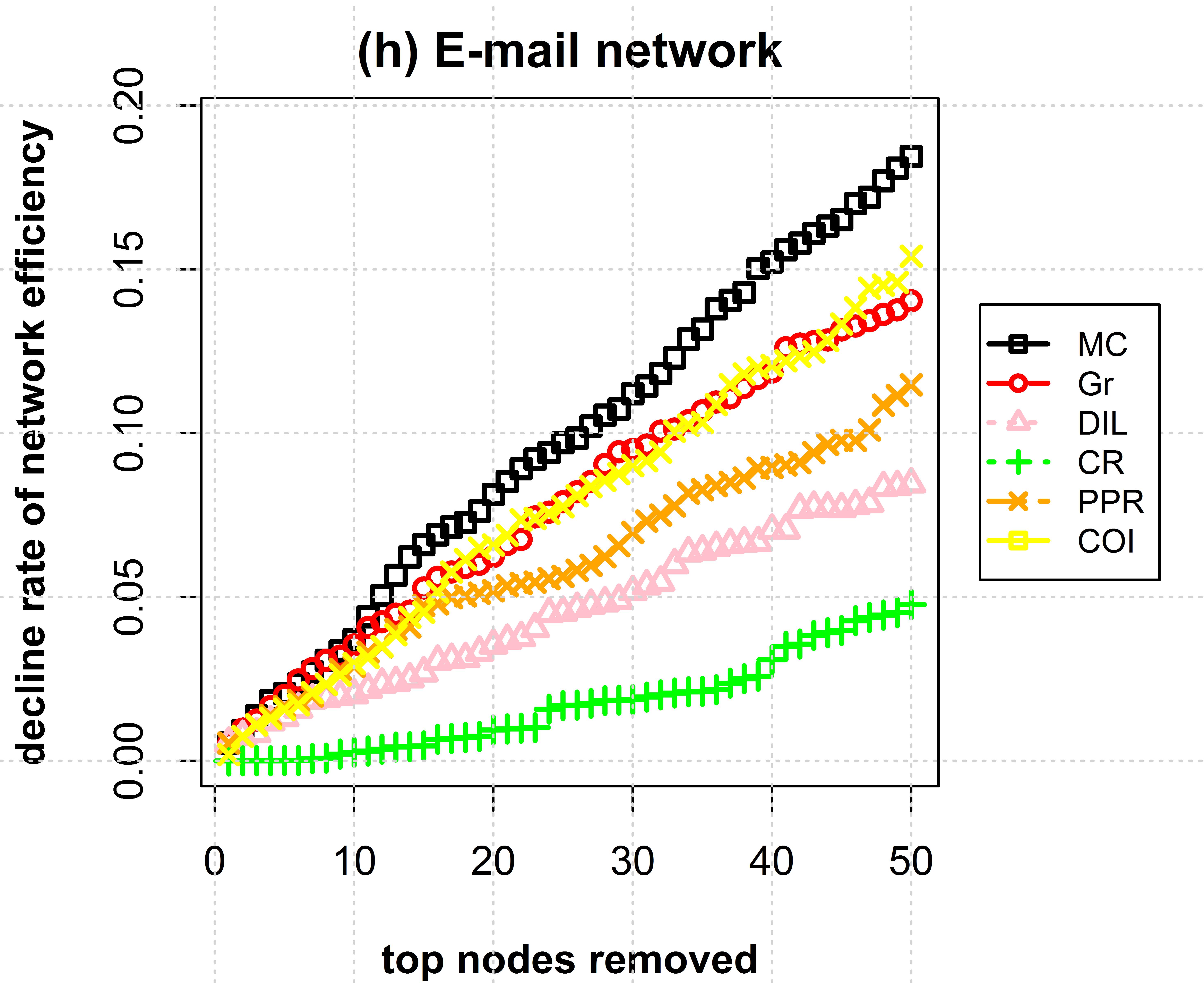}
\label{fig:8}}
\hfil
\caption{Decline rate of network efficiency for the eight real-world networks under study.}
\label{figure 1}
\end{figure*}

After we study the structural damage caused by the removal of important nodes, we move on to study the impact of deleting important nodes on network efficiency. Figure \ref{figure 1} shows the relationship between the decline rate of network efficiency and the number of nodes removed from the network. Two main observations can be made. First, we can see that the decline rate of network efficiency is rising with the increase of the number of nodes removed. Second, the proposed measure clearly performs the best compared to its alternatives, with ClusterRank giving the worst results in all the networks under study. 

For Dolphins network, the removal of node SN100 (identified by DIL) seems to cause more damage to the network than nodes Grin and Hook. This comforts the fact that SN100 is a key member in holding the group together. Gravity and DIL give good performances in all stages of the removal process, while Personalized PageRank fails in the last stages (after the removal of the $8^{th}$ node). In Les Misérables, removing Valjean (identified by M-Centrality) causes more damage than Gavroche. In the stages 2, 3 and 4 of the removal process, we notice that Marius and Enjolras are more important than Myriel. The fifth stage of the removal process rises the fact that Javert is clearly more central than Bossuet. This suggests that the most suitable ranking would be Valjean, Gavroche, Marius, Enjolras and Javert. For Game Of Thrones, the top 10 nodes identified by M-Centrality are clearly the most important ones. Gravity, DIL and Personalized PageRank give competitive results. Globally all centralities are very competitive with a slight advantage of the proposed method due to its consistent ranking as previously shown in Table \ref{table:3}. For the remaining networks, the performance gap between M-Centrality and its alternatives is consequent and more visible especially in the cases of Political blogs, US airport and E-mail networks. In Netscience and Paul Erdős, Collective Influence gives better results in the first network and is as competitive as M-Centrality in the second one. This is concordant with the results presented in Table \ref{table:10} and can be explained by the fact that in this type of networks, the importance of an author is closely related to the importance of authors with whom he collaborates. This topological property is captured perfectly by the definition of the Collective Influence centrality.

Being said, given the computational complexity, monotonicity, structural damage caused by the removal of key nodes and network efficiency scores, the proposed centrality gives better results than its alternatives, which comforts the previous ranking results presented in Section \ref{sec:1}. This suggests a more consistent and accurate ranking. We also notice that the highest vulnerability to the removal of important nodes identified by M-Centrality is registered for Game Of Thrones (90\%) and Les Misérable (87\%) while the lower one is for E-mail network (18\%). This reflects that some networks are more resilient to targeted attacks than others.

\begin{figure*}[!h]
\centering
\subfigure{
\includegraphics[width=2.3in]{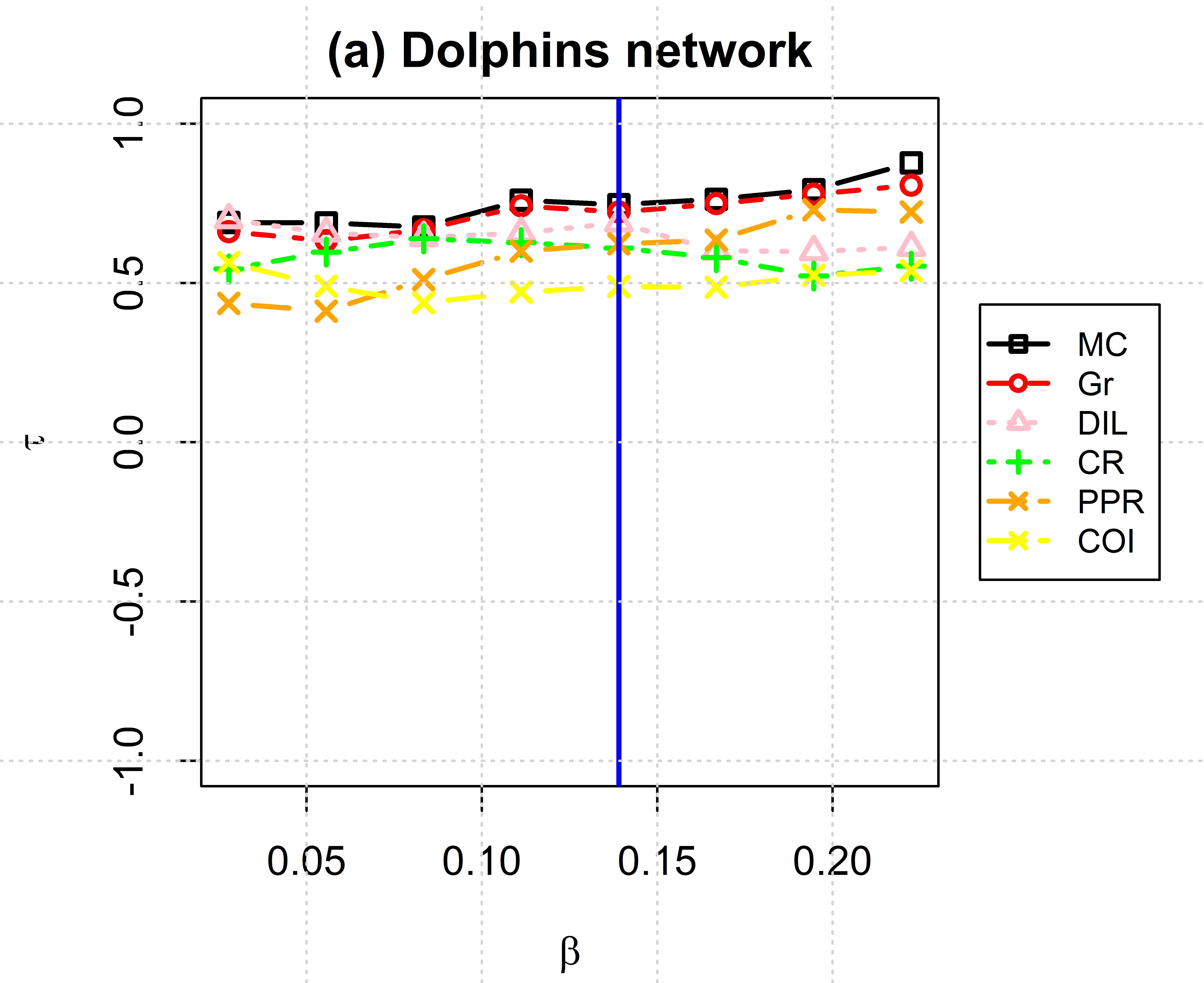}
\label{fig:1}}
\hfil
\subfigure{
\includegraphics[width=2.3in]{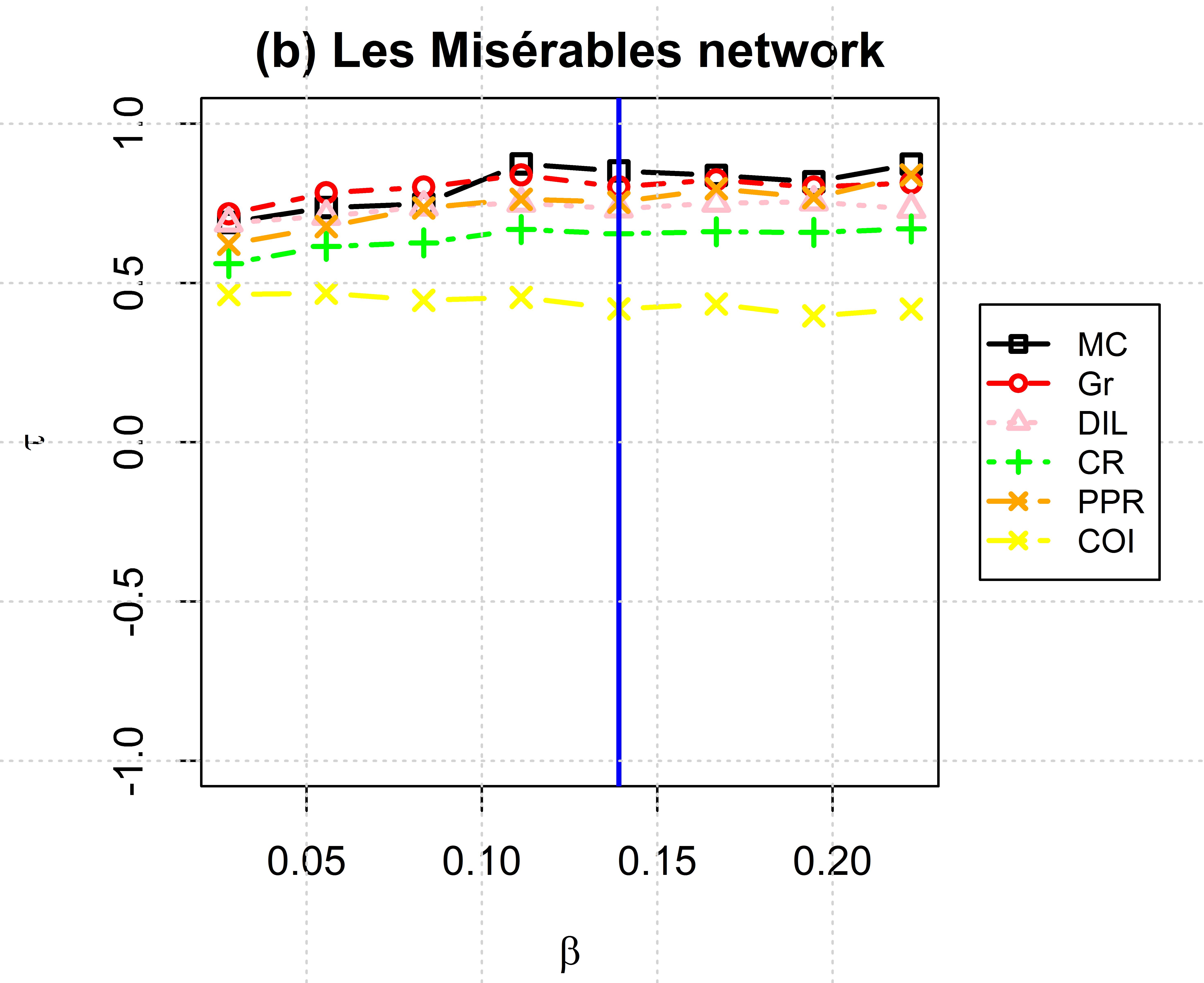}
\label{fig:2}}
\hfil
\subfigure{
\includegraphics[width=2.3in]{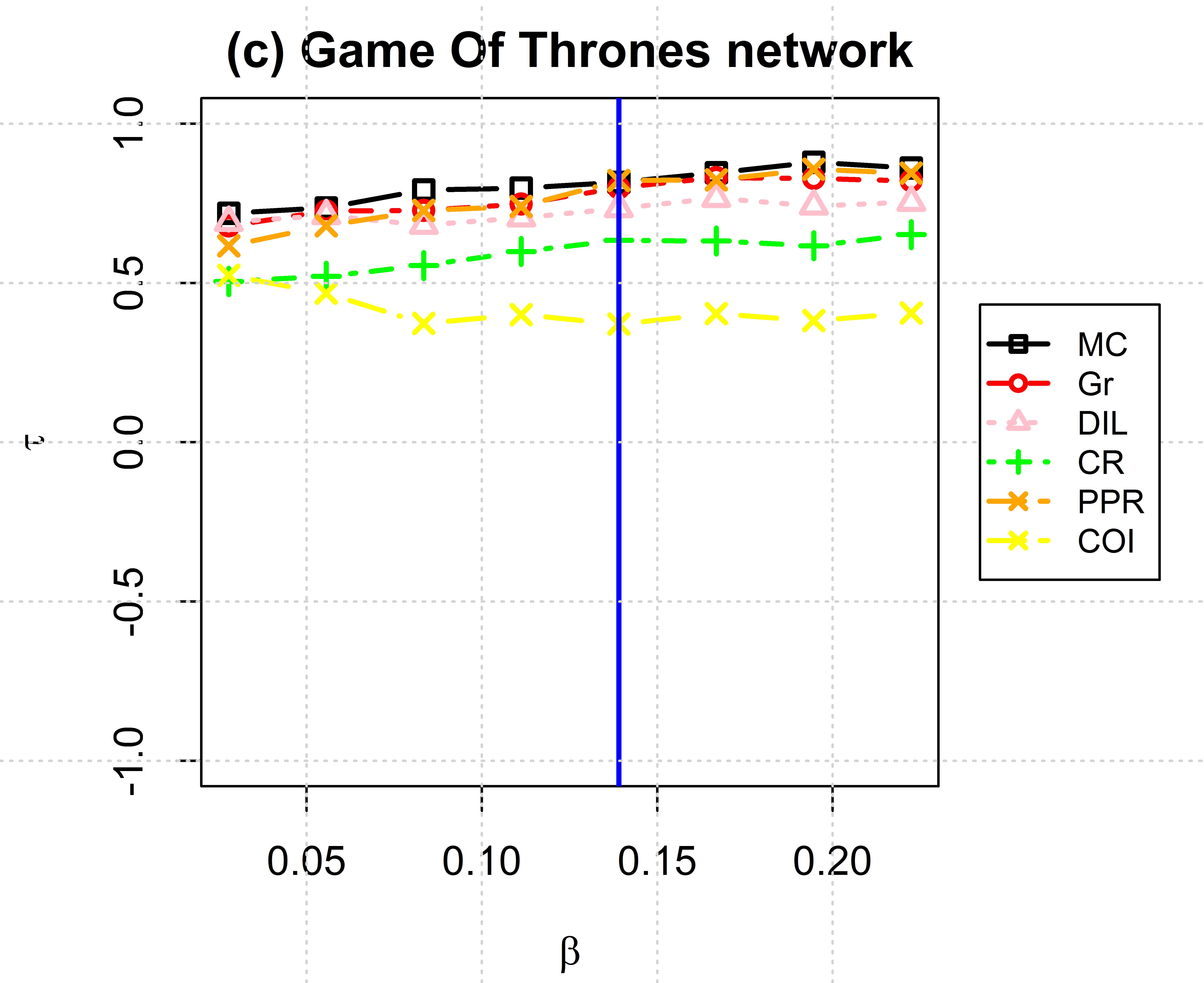}
\label{fig:3}}
\hfil
\subfigure{
\includegraphics[width=2.3in]{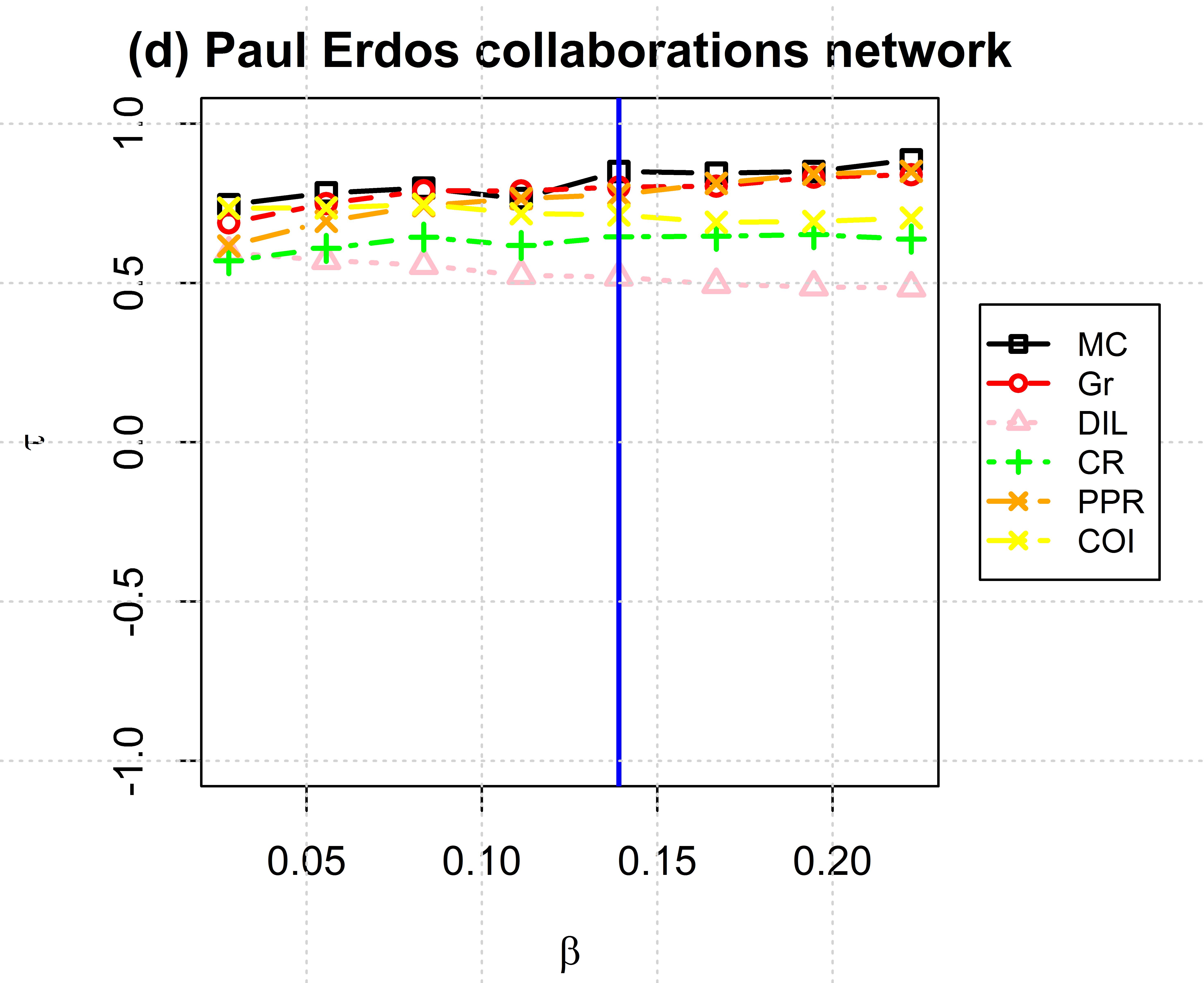}
\label{fig:4}}
\hfil
\subfigure{
\includegraphics[width=2.3in]{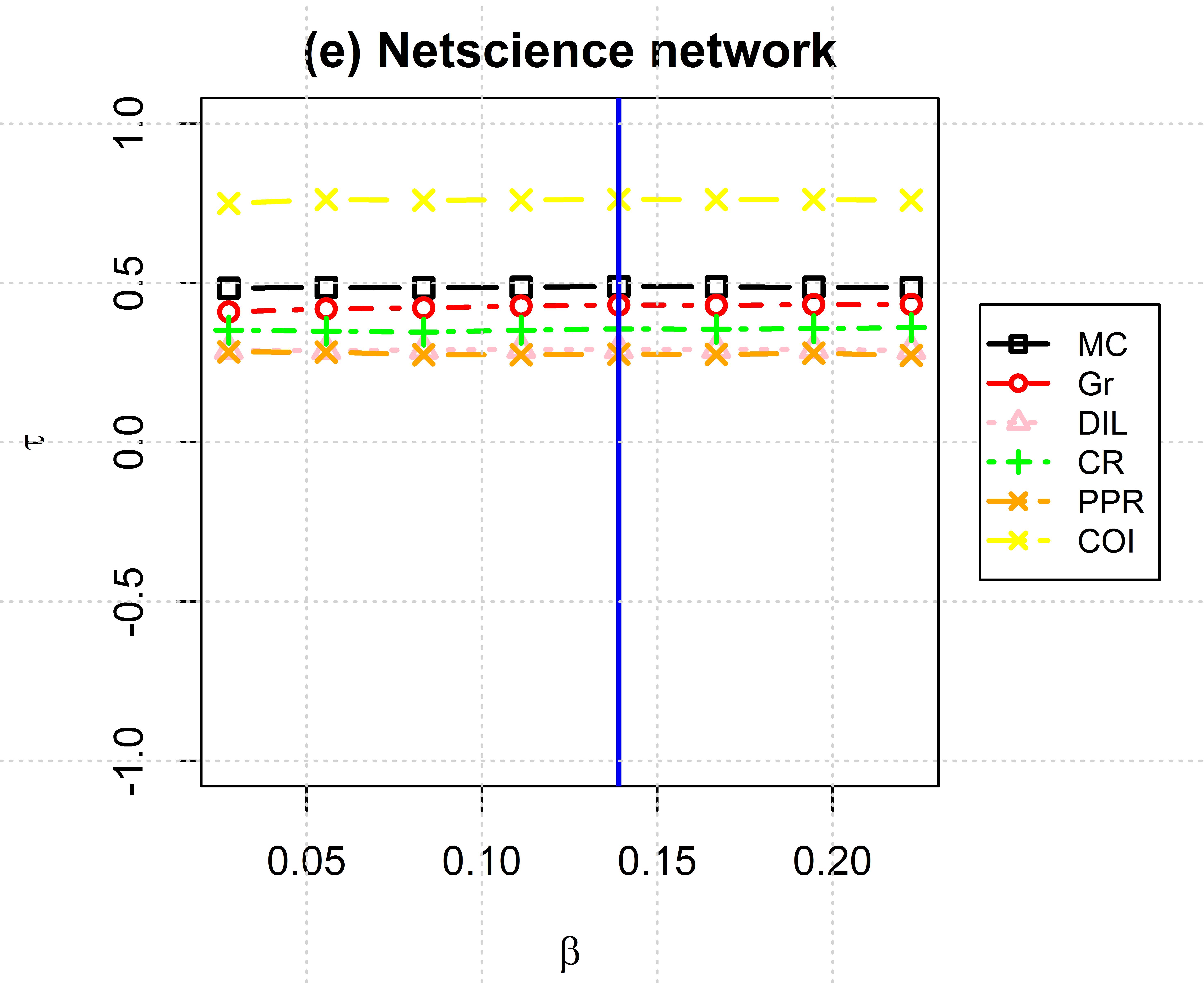}
\label{fig:5}}
\hfil
\subfigure{
\includegraphics[width=2.3in]{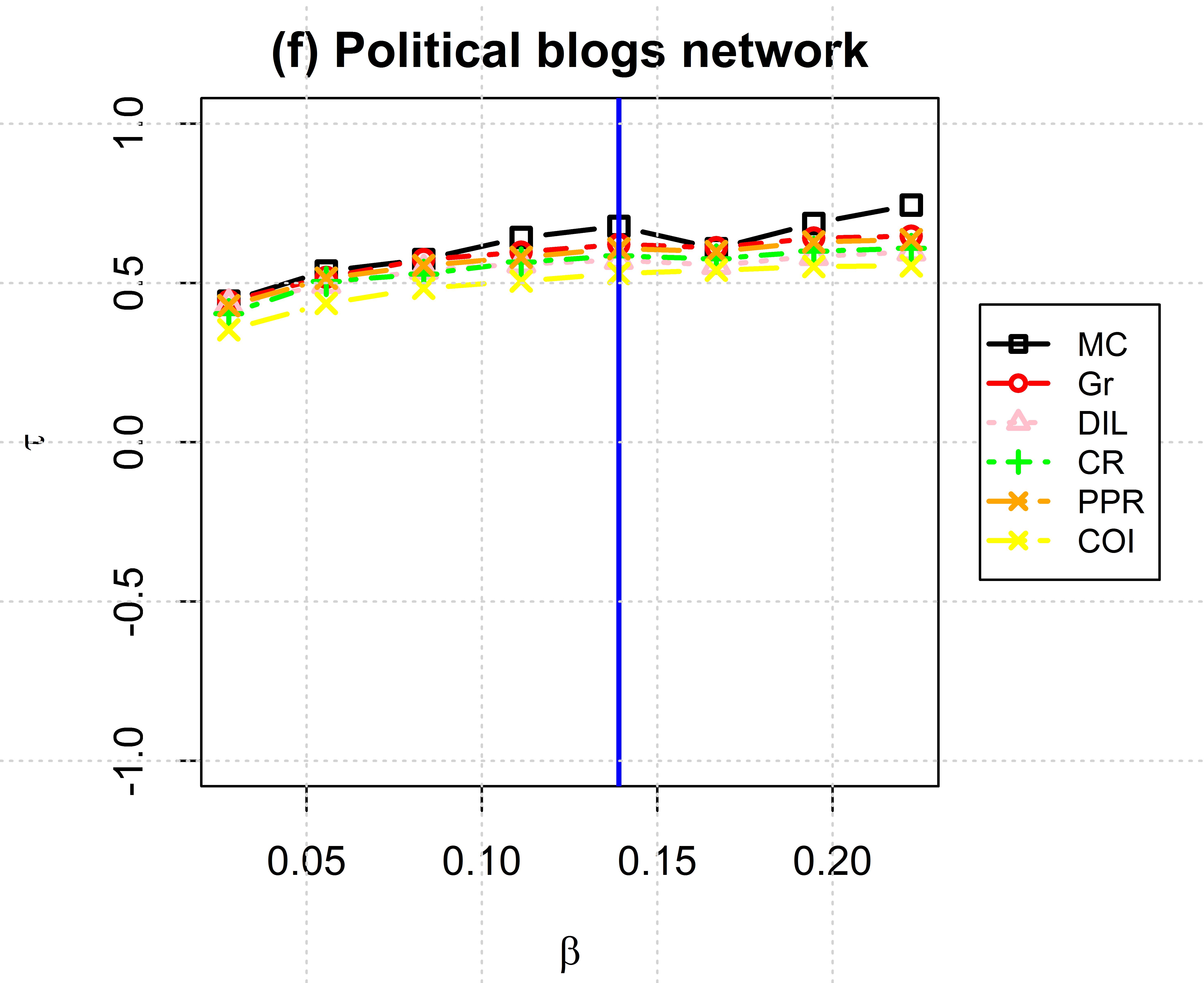}
\label{fig:6}}
\hfil
\subfigure{
\includegraphics[width=2.3in]{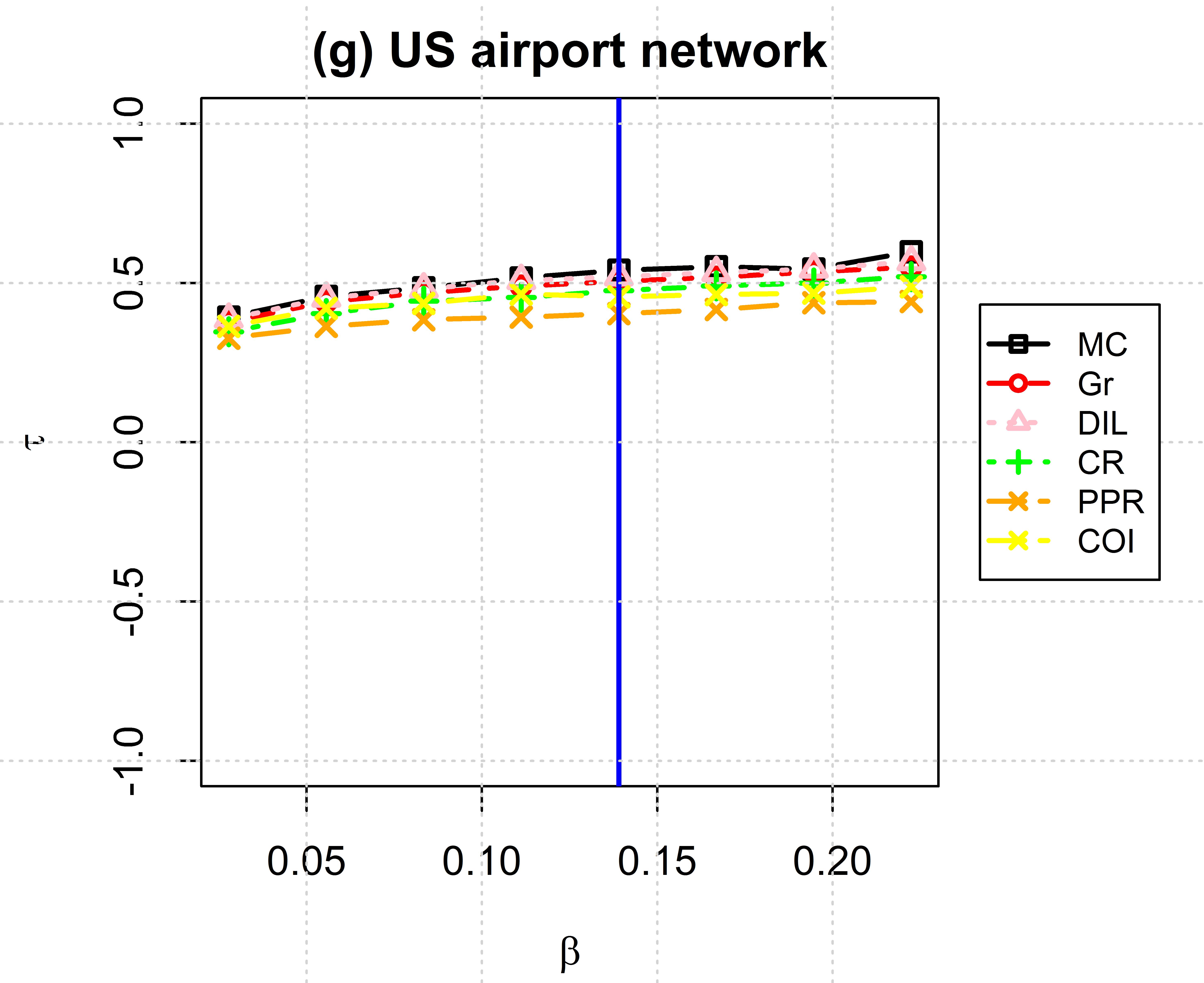}
\label{fig:7}}
\hfil
\subfigure{
\includegraphics[width=2.3in]{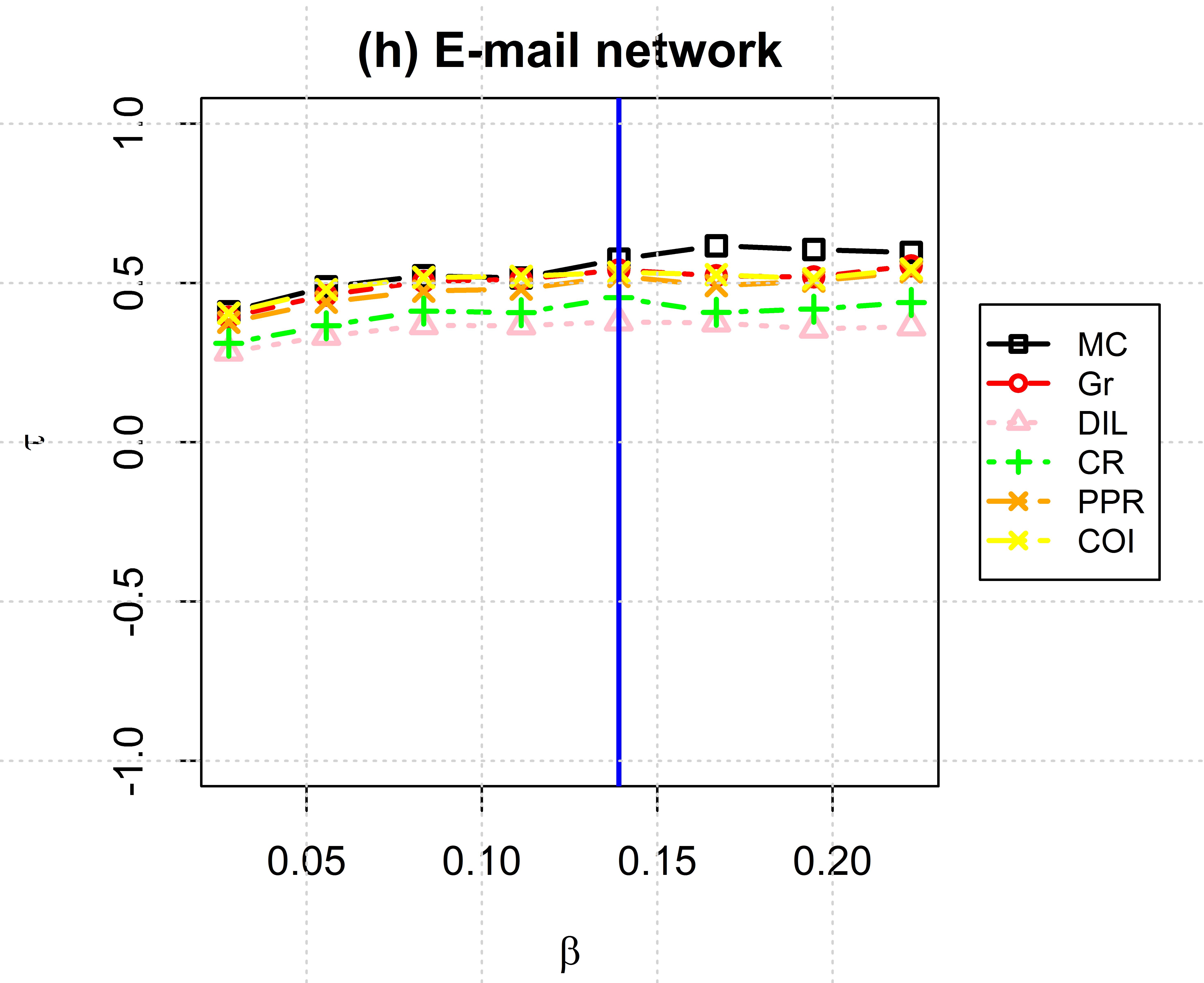}
\label{fig:8}}
\hfil
\caption{The accuracy of five centrality measures in evaluating the spreading influence of nodes according to the SIR model in the eight real-world networks, quantified by the Kendall’s Tau coefficients. The dot line correspond to the epidemic threshold $\beta_{th}$.}
\label{figure 2}
\end{figure*}

\subsection{Evaluation of nodes spreading capability with SIR model}

In addition to network vulnerability, we test the effectiveness of the methods by evaluating their Kendall tau correlation with the SIR model for different values of $\beta$.  A greater correlation indicates the greater accuracy of the centrality in quantifying the spreading capability of nodes.  Figure \ref{figure 2} shows the rank correlation results of different ranking methods on the previously studied networks. Note that M-Centrality is compared with the most recent and effective methods in detecting influential spreaders. For the real-world networks investigated, the first observation to make is that the correlation of different methods with the SIR epidemic model increases as the transmission rate increases. Second, for small values of $\beta$, the correlation is low. This is due to the fact that the epidemic is limited to the neighborhood of nodes and thus couldn't spread all over the network. Third, when $\beta$$>$$\beta_{th}$, we can observe that the variations in the transmission rate have in general little effect on the correlation values. This can be explained by the fact that when the transmission rate is larger than the epidemic threshold, the epidemic starting from any node can reach all the network nodes and the role of topology in the spreading will be overridden. Thus, distinguishing between nodes spreading capabilities become much harder. This is the case for all centralities except M-Centrality that seems to have the ability to capture these small variations in the transmission rate due to its local component $\Delta\mathfrak{D}$. This is an interesting feature of the proposed method that needs to be perfected in the future. We also can see that despite all centralities being very competitive, better results are achieved by M-Centrality and Gravity with a slight advantage for the first one, see Table \ref{apptable5} for the numerical results corresponding to Figure \ref{figure 2}. This is reasonable since they are both the improved methods of k-shell method. Although the M-Centrality evaluates the node influence from multiple attributes, the proposed method weight the position and neighborhood attributes by using the entropy method, leading to a better rank correlation and thus better performances than its alternatives. Overall, in the cases of the three small networks and Political blogs, the less competitive and inconsistent centralities are Collective Influence and ClusterRank. For Netscience, E-mail and Paul Erdős collaborations, it is DIL centrality that fails to deliver suitable results. These results confirm that the nodes located in the core of the network are more important when considering epidemic spreading.

The main problem of centralities is that they characterize the importance of nodes related to the topological features, which is perfect when we are considering the structural importance of nodes. But they may not perform well in processes involving transmission dynamics which consider more about the dynamics of nodes \cite{liu2016locating}. This can explain why in Figure \ref{figure 2} the correlation between the different centralities and the SIR model is not perfect. 

\subsection{Further Analysis}

We conclude the analysis by highlighting the common features of the M-Centrality revealed by the investigations on real-world networks. In order to measure a node influence, the M-Centrality combines a global information about its position in the network with the local information of the degree variation in its neighborhood. According to the weight value, more or less importance is given to these complementary aspects of a node influence, see Table \ref{table:11}. In any case, varying the value of the mixing proportion $\mu$ allows to tune the influence of the node. For $\mu =0$, its influence depends only on the topology of its neighborhood, while for $\mu =1$, it is related to its global position in the network. Between these two extremes, M-Centrality tends to adapt to both local and global topological properties of the network. So far, the empirical results demonstrate the effectiveness of the entropy weighted technique to estimate the optimum mixing proportion of the M-Centrality components. Moreover, small variations around the optimal weight value does not significantly alter the ranking of the nodes. 

\begin{table}[h!]
\renewcommand{\arraystretch}{1.4}
\caption{Entropy associated with the global ($E_{K_s}$) and local measure ($E_{\Delta\mathfrak{D}}$) of M-Centrality for the real-world networks.}
\label{table:11}
\centering
\resizebox{4in}{!}{
\begin{tabular}{|c||c  c  c c c c c c|}
    \hline
    Network  &Dolphins & Les Misérables& Game Of Thrones& Paul Erdős collaborations &  Netscience &   US airport& E-mail & Political blogs\\
    \hline\hline
  $E_{K_s}$    & 0.44& 0.33 & 0.36 &0.5 &0.5&0.67 &0.40&0.5 \\
  \hline
  $E_{\Delta\mathfrak{D}}$ &0.56 &0.67 & 0.64&0.5 & 0.5 & 0.33&0.60&0.5\\
   \hline
\end{tabular}
}
\end{table}

These results show that M-Centrality is able to extract the most relevant information from the network topology at both the global and local level in order to measure the influence of the nodes. The local influence as measured by the degree variation in the neighborhood, allows to distinguish between the multiple nodes that share the same position in the network according to the Coreness centrality. For the real-world networks under study, it appears that M-Centrality shows a high and positive correlation, especially with Gravity centrality. This is due to the fact that they are both a variant of Coreness centrality. However, the results show that the proposed measure performs better in term of quantifying node influence topologically and in transmission dynamics.

The proposed measure, M-Centrality, is more subtle in ranking nodes than Coreness centrality. This subtlety can be summarized in four major points: 

\begin{enumerate}
\item Its hybrid nature (combine global and local information) that leads to a more accurate ranking of nodes. To illustrate this behavior, let's consider nodes having the same degree and Coreness in these networks, and let's see if M-Centrality can discern differences between them in term of influence. Table \ref{table:12} reports centrality informations about different nodes that share the same degree centrality and Coreness values in the three well documented networks under study. Results shows that these nodes can be always distinguished by the M-centrality because the information about the degree variation in their neighborhood is always different. For example, node Topless and node SN4 of the Dolphins network are ranked similarly by the Coreness and degree centralities, while they have different influence according to M-Centrality  (Topless is considered more central than SN4). This is due to the fact that the degree variation in the neighborhood of Topless ($\Delta\mathfrak{D}_{Topless}$ = 6.28) is higher than the one of SN4 ($\Delta\mathfrak{D}_{SN4}$ = 5.70). So the distinction between nodes having the same Coreness and degree is made subtly thanks to $\Delta\mathfrak{D}$. We also notice that  M-Centrality is higher when the nodes are strategically positioned in the core of network.
\item Invariance to change of attribute weights as shown by the wide range of $\mu$ values for which ranking accuracy is conserved. 
\item The very high monotonicity scores achieved on all the networks studied. In fact the hybrid nature of the method and the entropy weighting strategy play a major role.
\item Presents more correlation with the SIR model in the case of epidemic spreading, which results in a better detection of influential spreaders.
\end{enumerate}

\begin{table}[h!]
\renewcommand{\arraystretch}{1}
\caption{M-Centrality, degree variation in the neighborhood ($\Delta\mathfrak{D}$), degree and Coreness centralities of some nodes sharing the same degree and Coreness values in the Dolphins, Les Misérables and Game Of Thrones networks.}
\label{table:12}
\centering
\resizebox{4in}{!}{
\begin{tabular}{|c|| c|  c c c c|}
    \hline
    Network & Nodes ID & M-Centrality& $\Delta\mathfrak{D}$ & Degree & Coreness\\
    \hline\hline
    Dolphins &Topless, SN4&5.25, 4.93  &6.28, 5.70& 11 & 4\\
    \cline{2-6} 
               & Mus, Notch & 2.12, 2.07& 1.41, 1.31& 3 & 3 \\
    \cline{2-6} 
               & TR120, TR88 &1.51, 1.37 &1.11, 0.85& 2 & 2 \\
    \cline{2-6} 
               & Whitetip, Zig& 0.93, 0.81&0.87, 0.66& 1 & 1\\
   \hline
    Les Misérables &Brujon, Dahlia& 4.40, 3.12&3.10, 1.18& 7 & 7\\
    \cline{2-6} 
               & Mme Magloire, Anzelma& 2.63, 2.52&2.44, 2.28 & 3 & 3\\
    \cline{2-6} 
               &Perpetue, Magnon&1.71, 1.70&1.57, 1.55& 2 & 2\\
    \cline{2-6} 
               & Gribier, Jondrette& 0.83, 0.66& 0.75,0.50& 1 & 1\\
   
    \hline
    Game Of Thrones &Bronn, Theon& 3.44, 3.42&3.12, 3.09& 4 & 4\\
    \cline{2-6} 
               &Viserys, Eddison& 2.69, 2.61&2.51, 2.40 & 3 & 3\\
      \cline{2-6} 
               &Jeyne, Chataya&1.88, 1.87&1.81, 1.80& 2 & 2\\
    \cline{2-6} 
               & Doran, Orell& 0.98, 0.97& 0.97,0.96& 1 & 1\\
   \hline          
\end{tabular}
}
\end{table}

\section{conclusion}
In this paper, a new centrality measure combining global information related to the position of the node in the network and local information linked to its neighborhood is proposed. The node position in the network is measured by its Coreness and the local information is quantified by the degree variation in its neighborhood. Rather than assigning the same importance to both components of the M-Centrality, a weighting strategy based on their respective entropy is applied in order to tune their relative importance to the network topology. An extensive comparative evaluation has been performed on eight real-world networks of different scales, and with existing as well as unknown knowledge about influential nodes. The experimental analysis provides strong evidence of the effectiveness of the M-Centrality as compared to recent and robust centrality measures such as Gravity, DIL, Personalized PageRank, ClusterRank and Collective Influence. Despite the fact that it shows strong correlation with Gravity centrality, M-Centrality gives better results in term of capturing topological and dynamical influence of nodes. It is therefore more subtle in measuring node influence. Moreover the proposed method exhibits a low complexity ($\mathcal{O}(n)$) which makes it suitable for large scale networks. This work can be extended in various directions. First of all, other variants of K-shell, including MDD and INK can be substituted to the position component of the measure. Further progress can be obtained by a finer evaluation of the degree variation in the neighborhood. Indeed, all the neighbors are given the same weight independently of their context. Finally, extensions to weighted and directed networks must be addressed.

\appendix

\section{Impact of the weight $\mu$ on M-Centrality}\label{regsol}

\setcounter{figure}{0}
\renewcommand\thefigure{\Alph{section}.\arabic{figure}}
\setcounter{table}{0}
\renewcommand\thetable{\Alph{section}.\arabic{table}}

Table \ref{apptable1} reports for Dolphins and Les Misérables networks the top 15 nodes sorted according to their M-Centrality values computed using various weights ranging from $\mu=0$ to $\mu=1$. In Dolphins, for $\mu$ = 0, the M-Centrality reduces M = $\Delta\mathfrak{D}$. The proposed measure identify nodes Grin, Trigger, Topless, Jet and Web as the 5 most important (See Table \ref{apptable1}). It also succeed in identifying SN100 in the top 15. When the weight $\mu$ increases, the rank of SN100 remains constant. For $\mu$ = 1, there is four groups of nodes. The biggest group with the highest coreness value  ($K_s=4$) contains 36 individuals including node SN100, Grin, Jet and Kringel that are considered among the most influential by $\Delta\mathfrak{D}$). The second one with $K_s=3$ contains 9 individuals. The third one with $K_s=2$ contains 8 individuals. The last one ($K_s=1$) is made of 9 individuals. These results confirm the inability of Coreness to distinguish accurately the nodes in a networks. For Les Misérables network, when $\mu$ = 0, the M-Centrality measures the degree variation in the neighborhood of the nodes (M = $\Delta\mathfrak{D}$). It ranks Valjean as the most central, followed by Gavroche, Marius, Javert, Fantine, Mr and Mme Thenardier and Cosette. Those are main characters of the novel. All the other nodes are not so central that the measure may suggest. At the other extreme, ($\mu= 1$) the M-Centrality reduces to the Coreness. Among the top fifteen nodes, 12 share a Coreness value of 9 and the 3 remaining have also a common coreness of 8. Note that there is some characters in this list such as Gavroche and Marius ($K_s=9$) that are ranked more important than characters such as Valjean, Javert and Thenardier ($K_s=8$). When the weight $\mu$ ranges between 0.25 to 0.75 the top 15 characters are always the same with the exception of Cosette that disappear from the list when $\mu=0.75$. Note that a high variation of $\mu$ is needed in order to observe some evolutions of the ranking. 

\begin{table}[h!]
\renewcommand{\arraystretch}{1}
\caption{First 15 nodes sorted by relevance according to M-Centrality, Coreness and $\Delta\mathfrak{D}$  in Dolphins and Les Misérables networks. The numbers between parenthesis correspond to node centrality values.}
\label{apptable1}
\centering
\resizebox{4in}{!}{
\begin{tabular}{|c|| c|  c  c c c c|}
\hline

    Network&Rank &  $\Delta\mathfrak{D}$($\mu = 0$) & M($\mu = 0.25$)& M($\mu = 0.50$)& M($\mu = 0.75$)   & $K_s$($\mu = 1$)    \\
    \hline\hline
\multirow{15}{*}{Dolphins}&1&        Grin   &    Grin    &   Grin    &   Grin    &   Beak (4)   \\
&2 &    Trigger   & Trigger   & Trigger &   Trigger& Beescratch  (4) \\
&3 &    Topless&    Topless &   Topless  &  Topless &      DN21  (4) \\
&4 &        Jet     &   Jet   &     Jet    &    Jet   &    DN63  (4) \\
&5 &        Web     &   Web    &    Web    &    Web   &  Double  (4) \\
&6 &        SN4       & SN4    &    SN4   &     SN4  &  Feather  (4) \\
&7 &      Scabs    &  Scabs   &   Scabs    &  Scabs     &  Fish  (4) \\
&8 &  Patchback & Patchback & Patchback & Patchback  & Gallatin (4)  \\
&9   &  Kringel   & Kringel   & Kringel    &Kringel       &Grin (4)  \\
&10   &    SN63    &   SN63  &     SN63   &    SN63 &  Haecksel   (4)\\
&11 &Beescratch &Beescratch& Beescratch& Beescratch &      Hook   (4)\\
&12   & Stripes   & Stripes  &  Stripes  &  Stripes       & Jet  (4) \\
&13   &   \cellcolor{blue}SN100   &   \cellcolor{blue}SN100   &   \cellcolor{blue}SN100  &    \cellcolor{blue}SN100  &   \cellcolor{blue} SN100 (4)  \\
&14 &  Gallatin  & Gallatin &  Gallatin &  Gallatin   &    Knit   (4)\\
&15  & Shmuddel &       SN9    &    SN9 &       SN9  &Kringel   (4)\\
  \hline\hline

\multirow{15}{*}{Les Misérables}&1  &        \cellcolor{blue} Valjean        &  \cellcolor{blue}Valjean  &   \cellcolor{blue} Valjean   & \cellcolor{blue}Valjean &  \cellcolor{red}Gavroche (9)  \\
&2  &        \cellcolor{red}Gavroche        & \cellcolor{red}Gavroche    & \cellcolor{red}Gavroche &  \cellcolor{red}Gavroche   &  \cellcolor{pink}Marius  (9) \\
&3  &          Myriel        &   Myriel  &     Myriel  &   \cellcolor{pink}Marius  &   Mabeuf   (9)\\
&4  &          \cellcolor{pink}Marius       &    \cellcolor{pink}Marius     &  \cellcolor{pink}Marius     &  \cellcolor{yellow}Javert &  Enjolras   (9)\\
&5  &          \cellcolor{yellow}Javert       &    J  \cellcolor{yellow}avert    &     \cellcolor{yellow}Javert &\cellcolor{green}Thenardier& Combeferre (9)  \\
&6  &          \cellcolor{violet} Fantine      &    \cellcolor{violet}Fantine   &   \cellcolor{violet}Fantine &  \cellcolor{violet} Fantine & Prouvaire   (9)\\
&7  &      \cellcolor{green}Thenardier     &  \cellcolor{green}Thenardier &  \cellcolor{green}Thenardier &  Enjolras   & Feuilly (9)  \\
&8  &         \cellcolor{olive}Cosette       &  Enjolras &    Enjolras  &  Bossuet &Courfeyrac  (9) \\
&9  &        Enjolras        &   \cellcolor{olive}Cosette   &   Bossuet& Courfeyrac &   Bahorel  (9) \\
&10& MlleGillenormand         & Bossuet      & \cellcolor{olive}Cosette    & Mabeuf&    Bossuet   (9)\\
&11 &    Gillenormand &   \cellcolor{orange}MmeThenardier &  Courfeyrac &    Myriel   &    Joly  (9) \\
&12 &    \cellcolor{orange}MmeThenardier    & Montparnasse    &   Mabeuf & Prouvaire & Grantaire  (9) \\
&13 &      Bamatabois &MlleGillenormand &Montparnasse  &  Bahorel   & \cellcolor{blue}Valjean  (8) \\
&14&          Bossuet     &   Gueulemer &   Prouvaire  &     Joly& \cellcolor{green}Thenardier    (8)\\
&15&     Montparnasse &           Babet  &  Gueulemer&  Grantaire   &    \cellcolor{yellow}Javert    (8)\\

  \hline
\end{tabular}
}
\end{table}

For the two remaining small scale networks, Game Of Thrones and Paul Erdős collaborations, the same previous observations can be made. Table\ref{apptable2} shows that $\mu$ can vary in a wide range without changing the top nodes positions. This is interesting because the estimate of $\mu$ can be very approximative without affecting too much the M-Centrality behavior.

\begin{table}[h!]
\renewcommand{\arraystretch}{1}
\caption{First 15 nodes sorted by relevance according to M-Centrality, Coreness and $\Delta\mathfrak{D}$  in Game Of Thrones and Paul Erdős collaborations networks. The numbers between parenthesis correspond to node centrality values.}
\label{apptable2}
\centering
\resizebox{5.5in}{!}{
\begin{tabular}{|c|| c|  c  c c c c|}
\hline

   Network& Rank &  $\Delta\mathfrak{D}$($\mu = 0$) & M($\mu = 0.25$)& M($\mu = 0.50$)& M($\mu = 0.75$)   & $K_s$($\mu = 1$)    \\
    \hline\hline
\multirow{15}{*}{Game Of Thrones}&1  &  \cellcolor{blue}Tyrion & \cellcolor{blue} Tyrion  & \cellcolor{blue}Tyrion & \cellcolor{blue} Tyrion&  \cellcolor{violet} Jaime  (7) \\
&2   &    \cellcolor{red}Jon   &    \cellcolor{red}Jon    & \cellcolor{red}  Jon    &  \cellcolor{red} Jon & Robert  (7)  \\
&3  & \cellcolor{pink}Daenerys &\cellcolor{pink}Daenerys &    \cellcolor{yellow}Sansa  &   \cellcolor{yellow}Sansa & \cellcolor{blue}Tyrion   (7)\\
&4  &  \cellcolor{yellow} Sansa &    \cellcolor{yellow}Sansa  &   Robb   &  Robb  & Tywin   (7)\\
&5  &    Robb  &   Robb&\cellcolor{pink} Daenerys&    Tywin &  \cellcolor{orange} Arya  (7)  \\
&6  &   Tywin&    Tywin  &  Tywin   & \cellcolor{violet}Jaime & \cellcolor{green}Cersei   (7)\\
&7  &    \cellcolor{violet}Jaime  &  \cellcolor{violet}Jaime   & \cellcolor{violet}Jaime& \cellcolor{pink}Daenerys & Gregor  (7) \\
&8  & Samwell&  Samwell & Samwell&  Catelyn& Joffrey   (7)\\
&9  &   Mance & Catelyn & Catelyn &  \cellcolor{green}Cersei&  Sandor   (7)\\
&10&  Catelyn&   \cellcolor{green}Cersei  & \cellcolor{green}Cersei &   \cellcolor{orange} Arya &Catelyn   (7) \\
&11&   \cellcolor{green}Cersei &   Mance &    \cellcolor{orange}Arya &  Robert  &  Robb   (7) \\
&12&     \cellcolor{orange}Arya   &  \cellcolor{orange}Arya  &  Mance & Joffrey &   \cellcolor{yellow}Sansa    (7)\\
&13&   Robert &  Robert  & Robert & Samwell &Stannis  (7)  \\
&14&  Joffrey&  Joffrey&  Joffrey & Stannis & Eddard    (7)\\
&15&     Bran &    Bran  &   Bran   &  Bran &   Bran    (6)\\

    \hline\hline

\multirow{15}{*}{Paul Erdős collaborations}&1   &        HARARY, FRANK    &   HARARY, FRANK   &    HARARY, FRANK   &HARARY, FRANK        &      \cellcolor{gray} ALON, NOGA    (9)\\
&2   &      POMERANCE, CARL  & \cellcolor{gray}GRAHAM, RONALD L.&   \cellcolor{gray}GRAHAM, RONALD L. &  \cellcolor{gray}GRAHAM, RONALD L.   &         BABAI, LASZLO   (9) \\
&3   &    \cellcolor{gray}GRAHAM, RONALD L. &    POMERANCE, CARL  &   POMERANCE, CARL &   TUZA, ZSOLT       &    BOLLOBAS, BELA   (9) \\
&4   &          TUZA, ZSOLT        & TUZA, ZSOLT     &    TUZA, ZSOLT    & \cellcolor{gray}RODL, VOJTECH     & BURR, STEFAN ANDRUS    (9)\\
&5   &        \cellcolor{gray}RODL, VOJTECH   &    \cellcolor{gray}RODL, VOJTECH  &     \cellcolor{gray}RODL, VOJTECH & POMERANCE, CARL    &   CHUNG, FAN RONG K.   (9) \\
&6   &         SOS, VERA T.     &   SOS, VERA T.    &    SOS, VERA T. &  SOS, VERA T.      &  FAUDREE, RALPH J.    (9)\\
&7   &           \cellcolor{gray}ALON, NOGA     &     \cellcolor{gray}ALON, NOGA   &       \cellcolor{gray}ALON, NOGA  &\cellcolor{gray}ALON, NOGA     &       FRANKL, PETER    (9)\\
&8   &     SPENCER, JOEL H.  &  SPENCER, JOEL H. &   SPENCER, JOEL H. &  SPENCER, JOEL H.      &     FUREDI, ZOLTAN   (9) \\
&9  &        HAJNAL, ANDRAS&      HAJNAL, ANDRAS &     HAJNAL, ANDRAS &   HAJNAL, ANDRAS   &      GOULD, RONALD J.   (9) \\
&10&         BOLLOBAS, BELA   &   BOLLOBAS, BELA  &    BOLLOBAS, BELA  &  BOLLOBAS, BELA   &     \cellcolor{gray}GRAHAM, RONALD L.   (9) \\
&11 &           PACH, JANOS      &   PACH, JANOS   &      PACH, JANOS  &  PACH, JANOS      &    GYARFAS, ANDRAS   (9) \\
&12  &     STRAUS, ERNST G.  &  STRAUS, ERNST G.   & STRAUS, ERNST G.    &KLEITMAN, DANIEL J.       &    HAJNAL, ANDRAS    (9)\\
&13  &  KLEITMAN, DANIEL J. &KLEITMAN, DANIEL J. &KLEITMAN, DANIEL J. &   STRAUS, ERNST G.        &    HARARY, FRANK    (9)\\
&14  &      CHARTRAND, GARY   &  CHARTRAND, GARY  &   CHARTRAND, GARY  &  CHARTRAND, GARY   & JACOBSON, MICHAEL S..  (9)  \\
&15  &   CHUNG, FAN RONG K. & CHUNG, FAN RONG K.&  CHUNG, FAN RONG K.  &  CHUNG, FAN RONG K.  &            LEHEL, JENO   (9) \\

  \hline
\end{tabular}
}
\end{table}

For the large scale networks under study, Tables \ref{apptable3} and \ref{apptable4} illustrate the M-Centrality evolution of the nodes for various values of the weight. The main observation is that the proposed measure behavior resembles the one previously observed. This shows that M-Centrality behavior is not affected by the network scale. As a main conclusion, the proposed method is robust as it is not affected by changes in weights attributes and network size.

\begin{table}[h!]
\renewcommand{\arraystretch}{1}
\caption{First 15 nodes sorted by relevance according to M-Centrality, Coreness and $\Delta\mathfrak{D}$  in Netscience and US airport networks. The numbers between parenthesis correspond to node centrality values.}
\label{apptable3}
\centering
\resizebox{5in}{!}{
\begin{tabular}{|c|| c | c  c c c c|}
    \hline

    Network&Rank &  $\Delta\mathfrak{D}$($\mu = 0$) & M($\mu = 0.25$)& M($\mu = 0.50$)& M($\mu = 0.75$)   & $K_s$($\mu = 1$)    \\
    \hline\hline

\multirow{15}{*}{Netscience}&1    &    BARABASI, A    &   BARABASI, A  &    BARABASI, A    &  BARABASI, A    &   GIOT, L (19)\\
&2    &      NEWMAN, M    &     NEWMAN, M  &      NEWMAN, M    &    NEWMAN, M   &   UETZ, P  (19)\\
&3    &       JEONG, H       &   JEONG, H      &   JEONG, H      &   JEONG, H            &    CAGNEY, G  (19)\\
&4    &       YOUNG, M      &    YOUNG, M     &    YOUNG, M       &  YOUNG, M     &     MANSFIELD, T  (19)\\
&5   &   BOCCALETTI, S   &  BOCCALETTI, S  &  BOCCALETTI, S      &    UETZ, P  &       JUDSON, R  (19)\\
&6   &       OLTVAI, Z      &   OLTVAI, Z &       OLTVAI, Z     &   CAGNEY, G       &        KNIGHT, J  (19)\\
&7   &         SOLE, R         &  SOLE, R   &       SOLE, R  &   MANSFIELD, T       &        LOCKSHON, D  (19)\\
&8   &         ALON, U       &    ALON, U        &  ALON, U  &      OLTVAI, Z      &        NARAYAN, V  (19)\\
&9   &  DIAZGUILERA, A &   DIAZGUILERA, A   &     KURTHS, J  &        GIOT, L &     SRINIVASAN, M  (19)\\
&10 &        KURTHS, J   &      KURTHS, J  & DIAZGUILERA, A        &JUDSON, R &        POCHART, P  (19)\\
&11 &        LATORA, V   &      LATORA, V   &     LATORA, V   &     KNIGHT, J     &     QURESHIEMILI, A  (19)\\
&12 &            HU, G       &      HU, G        &    HU, G     & LOCKSHON, D        &      LI, Y  (19)\\
&13  &        KAHNG, B   &       KAHNG, B      &    MUTH, S &      NARAYAN, V     & GODWIN, B  (19)\\
&14  &     STAUFFER, D  &         MUTH, S  &      VICSEK, T &   SRINIVASAN, M  &  CONOVER, D  (19)\\
&15  &         MUTH, S     &  STAUFFER, D    &     KAHNG, B   &    POCHART, P &   KALBFLEISCH, T  (19)\\

  \hline\hline

\multirow{15}{*}{US airport}&1  & 766 & 766 & 766  &766  &766 (64)\\
&2 &  \cellcolor{gray}114 &\cellcolor{gray} 114 & \cellcolor{gray}114 & \cellcolor{gray}114 &  \cellcolor{gray}114 (64)\\
&3 &  877&  877 & 877 & 877 & 877 (64)\\
&4 &  709 & 709  &709  &709 & 709 (64)\\
&5 & \cellcolor{gray}1200& \cellcolor{gray}1200 &\cellcolor{gray}1200 &\cellcolor{gray}1200 & \cellcolor{gray}1200 (64)\\
&6 & 1016& 1016 &1016 &1016 & 1016 (64)\\
&7 &  \cellcolor{gray}500&  \cellcolor{gray}500 & \cellcolor{gray}500&  \cellcolor{gray}500& \cellcolor{gray} 500 (64)\\
&8 &  389&  389  &389&  389 & 389 (64)\\
&9 &  711&  711  &711&  711 &711 (64)\\
&10& \cellcolor{gray}1068& \cellcolor{gray}1068& \cellcolor{gray}1068 &\cellcolor{gray}1068  &\cellcolor{gray}1068 (64)\\
&11 & \cellcolor{gray}391 & \cellcolor{gray}391& \cellcolor{gray} 391  &\cellcolor{gray}391 & \cellcolor{gray}391 (64)\\
&12 & 215 & 215&  215  &215 & 215 (64)\\
&13 & \cellcolor{gray}435&  \cellcolor{gray}435&  \cellcolor{gray}435  &\cellcolor{gray}435&  \cellcolor{gray}435 (64)\\
&14 & 875&  875 & 875  &875 & 875 (64)\\
&15 & 685&  505 & 685  &685 & 32 (64)\\
\hline
\end{tabular}
}
\end{table}

\begin{table}[h!]
\renewcommand{\arraystretch}{1}
\caption{First 15 nodes sorted by relevance according to M-Centrality, Coreness and $\Delta\mathfrak{D}$  in E-mail and US Political blogs networks. The numbers between parenthesis correspond to node centrality values.}
\label{apptable4}
\centering
\resizebox{6in}{!}{
\begin{tabular}{|c|| c | c  c c c c|}
    \hline

   Network& Rank &  $\Delta\mathfrak{D}$($\mu = 0$) & M($\mu = 0.25$)& M($\mu = 0.50$)& M($\mu = 0.75$)   & $K_s$($\mu = 1$)    \\
    \hline\hline
\multirow{15}{*}{E-mail network}&1 & \cellcolor{gray}105 &\cellcolor{gray}105& \cellcolor{gray}105&\cellcolor{gray} 105& 299 (11)\\
&2 &  23  &23 & 23 & 23 &389 (11)\\
&3 & 333 &333 &333 &333 &434 (11)\\
&4 &  41  &41 & 41 & 41 &552 (11)\\
&5 &  \cellcolor{gray}16&\cellcolor{gray}  16&  \cellcolor{gray}16  &\cellcolor{gray}16& 571 (11)\\
&6 &  \cellcolor{gray}42&  \cellcolor{gray}42  &\cellcolor{gray}42 & \cellcolor{gray}42 &726 (11)\\
&7 & 233 &233 &233& 233& 756 (11)\\
&8 &  24 & 24 & 24  &24 &788 (11)\\
&9  & 14 & 14  &14  &14 &885 (11)\\
&10& \cellcolor{gray}196& \cellcolor{gray}196 &\cellcolor{gray}196 &\cellcolor{gray}196& 886 (11)\\
&11&  21 & 21&  21&  21 &887 (11)\\
&12& 355 &355 &578 &578 &888 (11)\\
&13 &578& 578 &355& 135 &  1 (10)\\
&14 &135& 135 &135 &355 &  2 (10)\\
&15 & 76 & 76  &76 & 76&   3 (10)\\

  \hline\hline

\multirow{15}{*}{US Political blogs}&1  &                     blogsforbush.com  & blogsforbush.com&      blogsforbush.com   &  blogsforbush.com      & aintnobaddude.com (36)\\
&2  &                         \cellcolor{gray}dailykos.com   &   \cellcolor{gray}dailykos.com         &   \cellcolor{gray}dailykos.com     &    \cellcolor{gray}dailykos.com     &   \cellcolor{gray}dailykos.com (36)\\
&3  &                     drudgereport.com&   drudgereport.com      & drudgereport.com&         drudgereport.com   &    \cellcolor{gray}atrios.blogspot.com (36)\\
&4  &                     \cellcolor{gray} instapundit.com  &  \cellcolor{gray} instapundit.com       &   \cellcolor{gray}instapundit.com     &   \cellcolor{gray} instapundit.com   &   atrios.blogspot.com/ (36)\\
&5  &               \cellcolor{gray} talkingpointsmemo.com &   \cellcolor{gray} talkingpointsmemo.com    & \cellcolor{gray}  talkingpointsmemo.com &     \cellcolor{gray} talkingpointsmemo.com   &     \cellcolor{gray}  talkingpointsmemo.com (36)\\
&6  &                 \cellcolor{gray} atrios.blogspot.com   &   \cellcolor{gray}atrios.blogspot.com      &   \cellcolor{gray}   atrios.blogspot.com   & \cellcolor{gray}   atrios.blogspot.com    &    blog.dccc.org (36)\\
&7  &                    powerlineblog.com   &   powerlineblog.com       &        powerlineblog.com   &     powerlineblog.com   &    bodyandsoul.typepad.com (36)\\
&8    &                 michellemalkin.com   &   michellemalkin.com      &     michellemalkin.com     &      michellemalkin.com &    busybusybusy.com (36)\\
&9   &                   truthlaidbear.com  &    truthlaidbear.com    &      truthlaidbear.com     &     truthlaidbear.com   &   corrente.blogspot.com (36)\\
&10  &               \cellcolor{gray}washingtonmonthly.com  &   \cellcolor{gray} washingtonmonthly.com     &\cellcolor{gray}   washingtonmonthly.com         &  \cellcolor{gray}washingtonmonthly.com    & \cellcolor{gray} washingtonmonthly.com (36)\\
&11   &    littlegreenfootballs.com/weblog &   littlegreenfootballs.com/weblog  &      littlegreenfootballs.com/weblog    & littlegreenfootballs.com/weblog     & anoldsoul.blogspot.com(36)\\
&12    &                   wizbangblog.com  &     wizbangblog.com     &    wizbangblog.com  &   wizbangblog.com   &  democrats.org/blog (36)\\
&13    &                    hughhewitt.com    &     hughhewitt.com  &   hughhewitt.com   &  hughhewitt.com   &  digbysblog.blogspot.com (36)\\
&14   &                       juancole.com     &     juancole.com       & juancole.com           &  juancole.com       & juancole.com (36)\\
&15  &                   lashawnbarber.com    &   lashawnbarber.com&    lashawnbarber.com       &  lashawnbarber.com    &  elayneriggs.blogspot.com (36)\\
\hline
\end{tabular}
}
\end{table}

\section{Correlation in high ranks between M-Centrality and its alternatives}\label{regsol2}

The rank-biased overlap (RBO) measure is used to compare the overlap of the two rankings at incrementally increasing depths \cite{webber2010similarity}. This measure examines the accuracy of the ranking list paying more attention to high ranks on the list by considering weights for different ranks and assigning greater weights to high ranks. The value of RBO between two ranking lists X and Y is calculated using Eq. \ref{eq:equation44}.

\begin{equation}
RBO(X, Y, p)= (1 - p) \sum_{d=1}^{n}p^{d - 1}A(X, Y, d),  RBO \in [0, 1]
\label{eq:equation44}
\end{equation}

where A(X,Y,d) is the value of overlap between two ranking lists X and Y up to rank d calculated by Eq. \ref{eq:equation55}, n is the number of distinct ranks on the ranking list, p is a tunable parameter in (0, 1) range, that models the user’s persistence (at each depth down the two lists, the user has probability p of continuing to the next rank, and inversely probability 1 - p of deciding to stop).A high value of RBO reflects a high correlation in high ranks between the two ranking lists. 

\begin{equation}
A(X, Y, d)=  \frac{X_{1:d} \cap X_{1:d}}{X_{1:d} \cup X_{1:d}},
\label{eq:equation55}
\end{equation}

where $X_{1:d}$ (resp. $Y_{1:d}$) represents the elements present in ranks 1 to d of list X (resp. Y). 

In this work, we investigate the value of RBO between the ranking list produced by the different centralities and the M-Centrality ranking list. The value of p varies from 0.4 to 0.9, and the value of RBO is calculated in consequence. 

\begin{figure*}[!h]
\centering
\subfigure{
\includegraphics[width=2.3in]{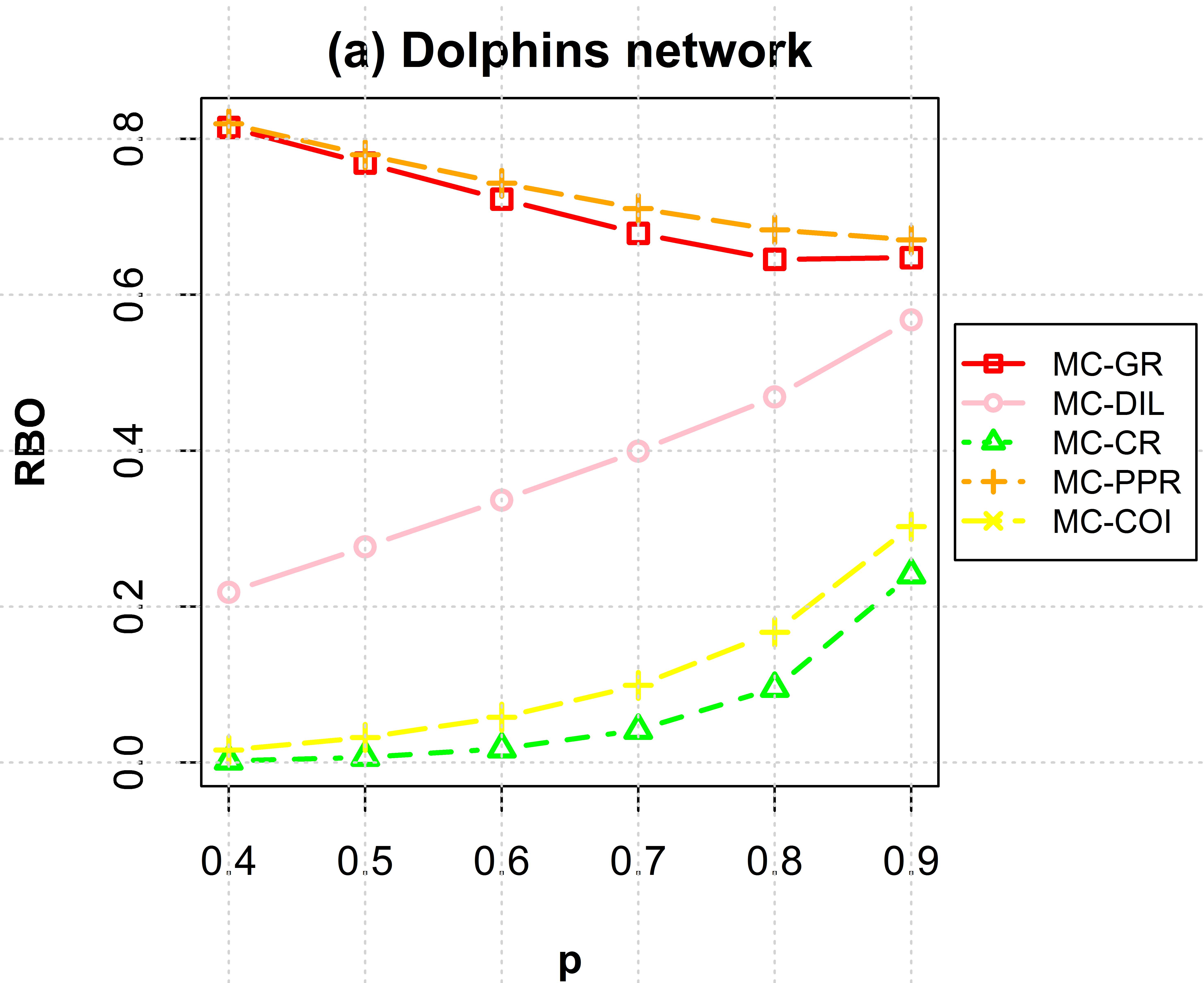}
\label{appfig:1}}
\hfil
\subfigure{
\includegraphics[width=2.3in]{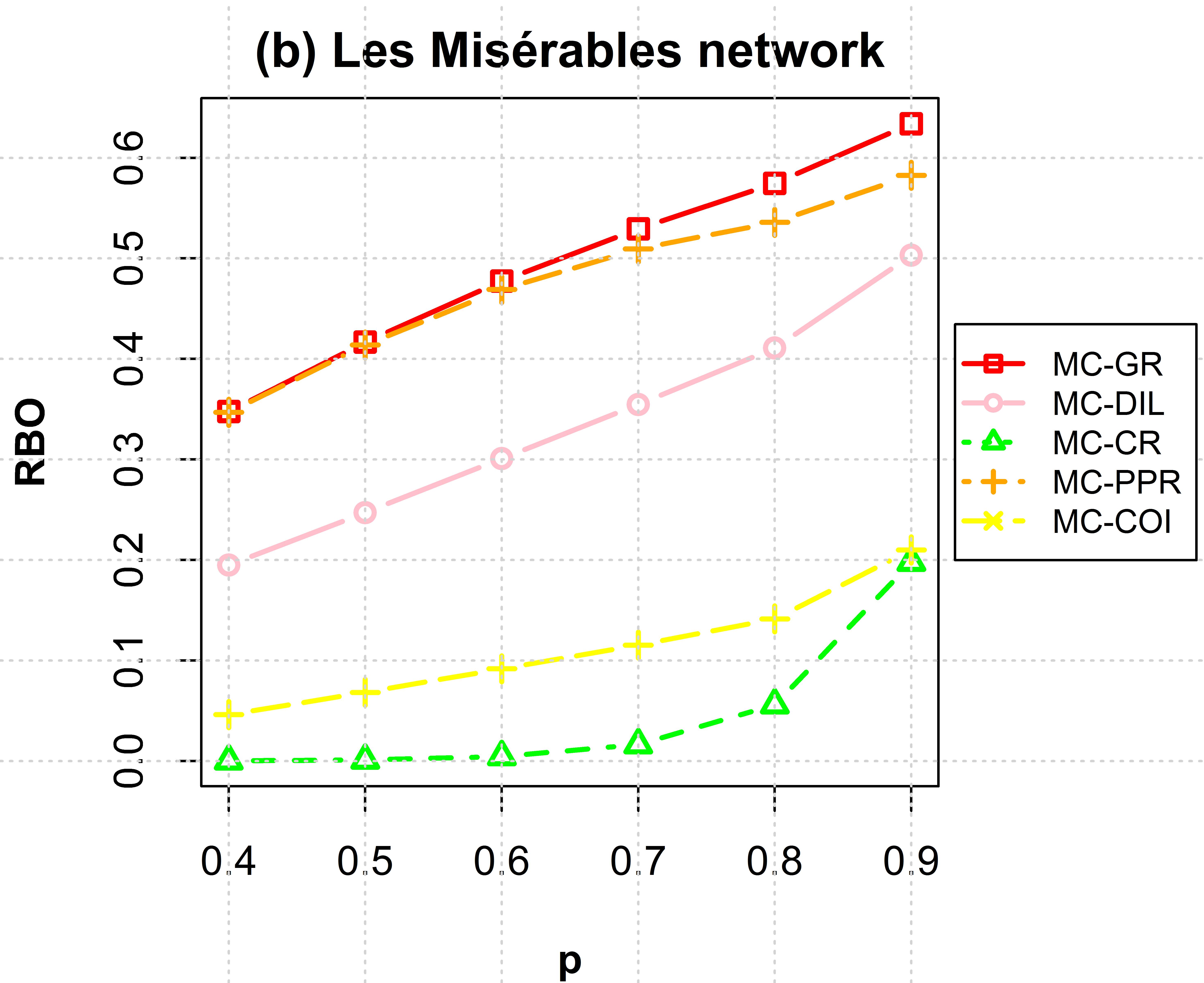}
\label{appfig:2}}
\hfil
\subfigure{
\includegraphics[width=2.3in]{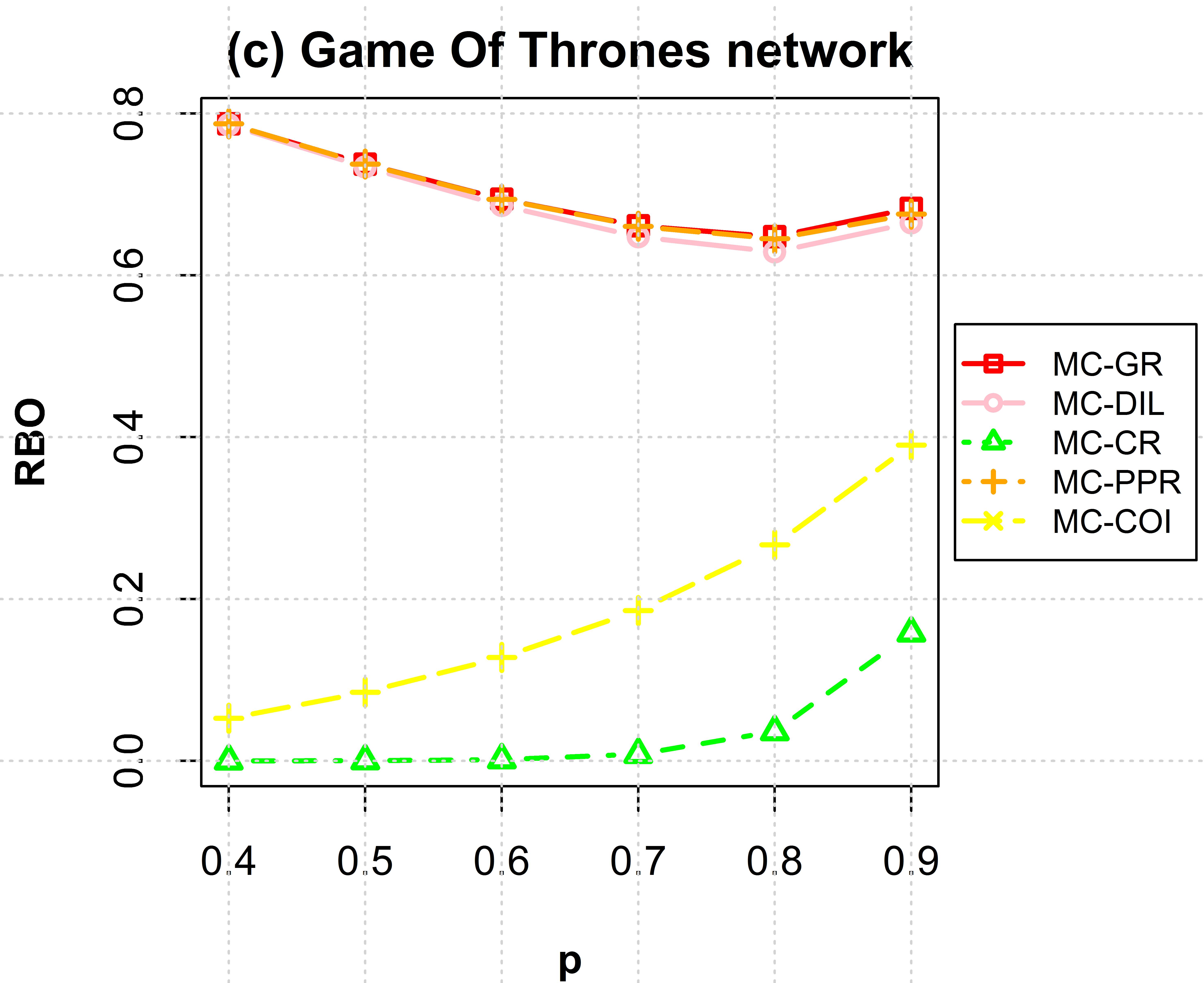}
\label{appfig:3}}
\hfil
\subfigure{
\includegraphics[width=2.3in]{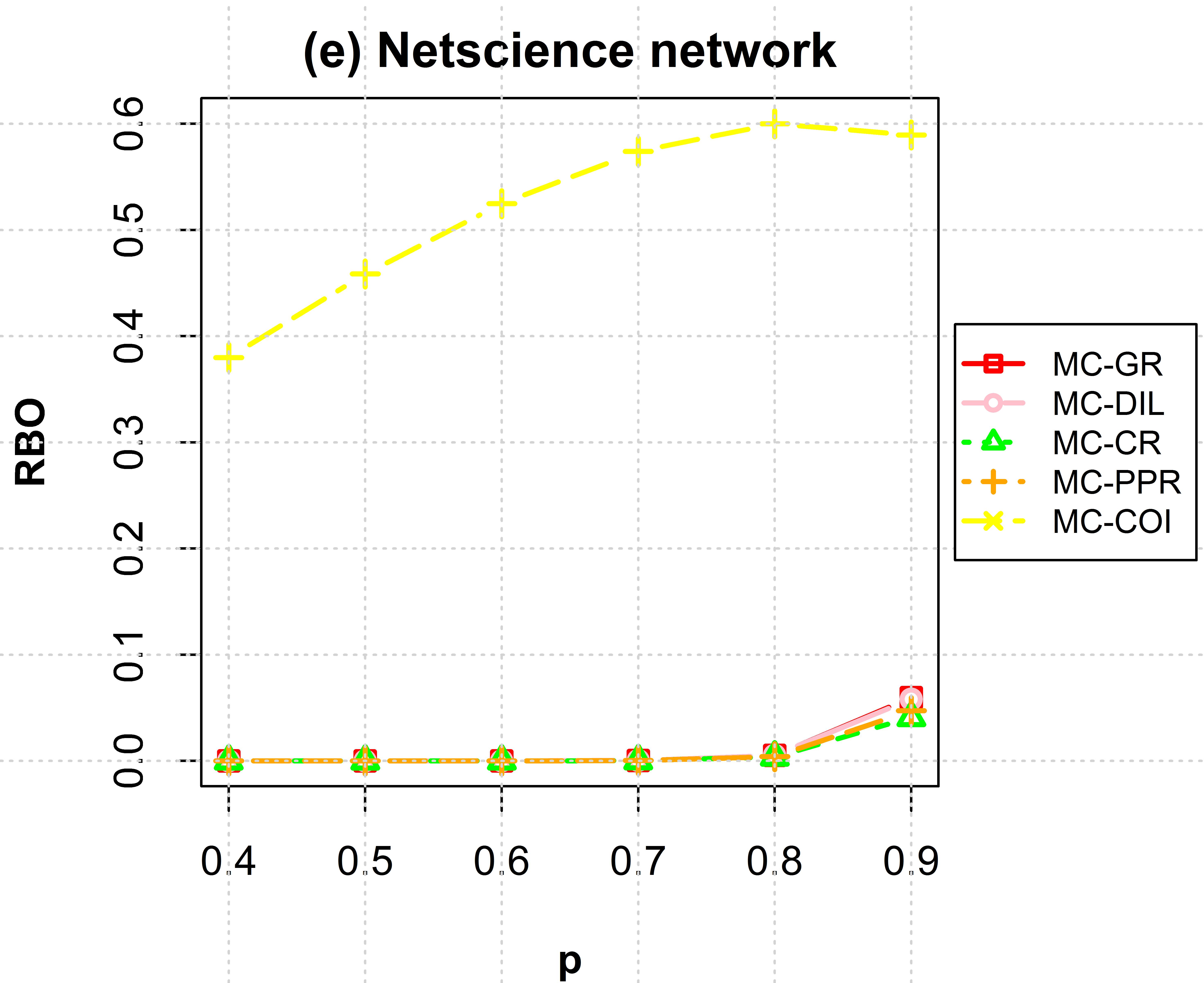}
\label{appfig:4}}
\hfil
\subfigure{
\includegraphics[width=2.3in]{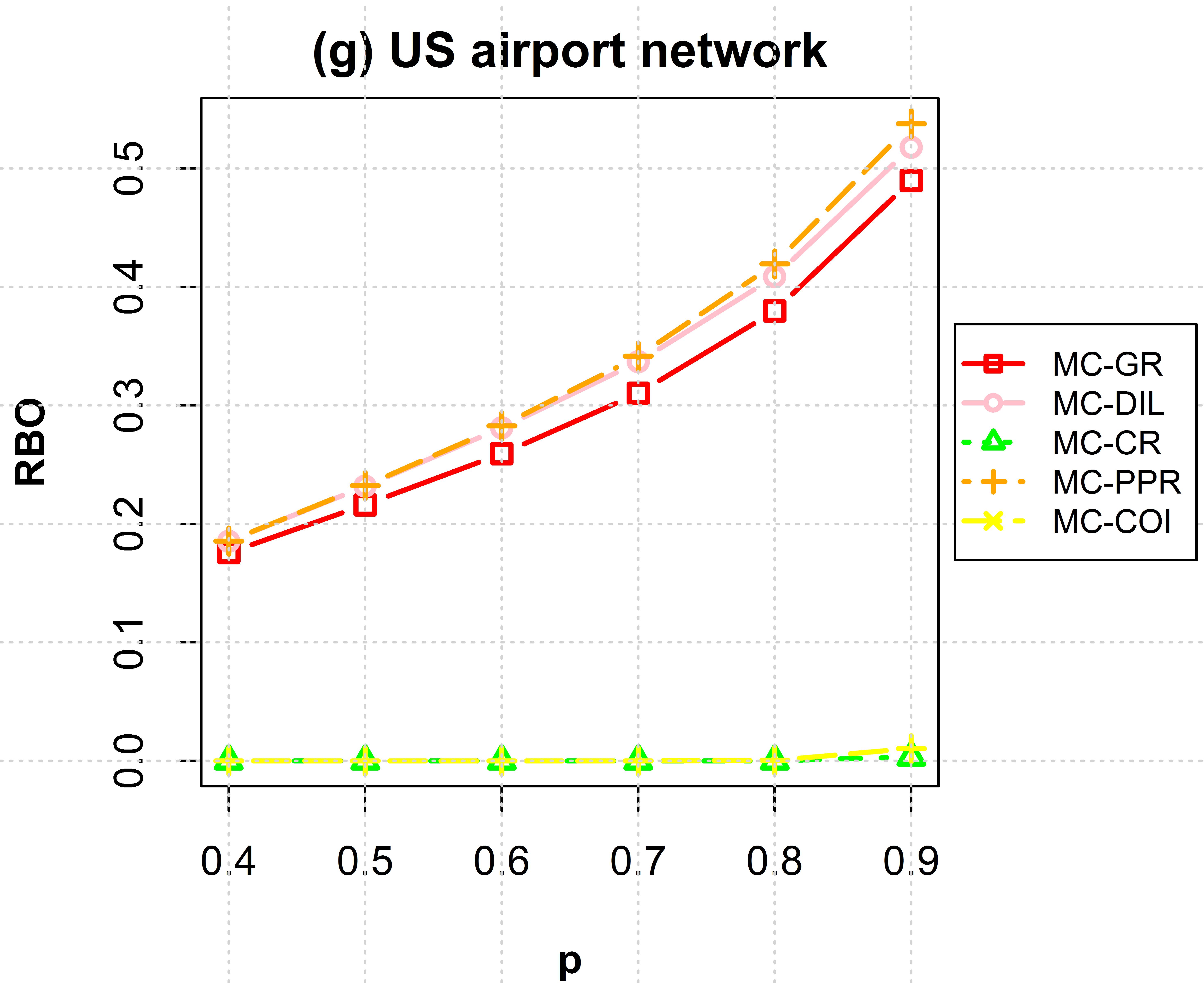}
\label{appfig:5}}
\hfil
\subfigure{
\includegraphics[width=2.3in]{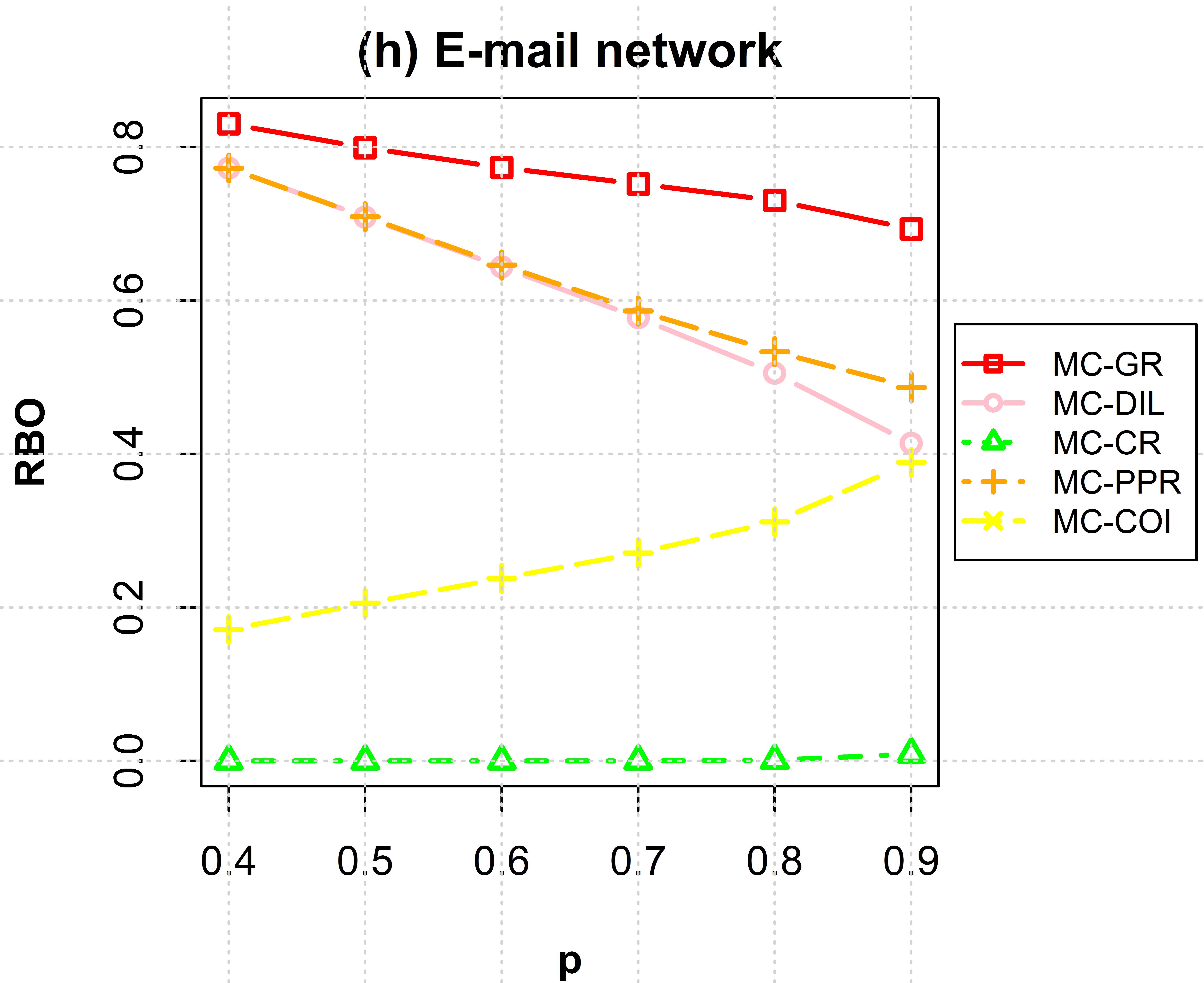}
\label{appfig:6}}
\hfil
\subfigure{
\includegraphics[width=2.3in]{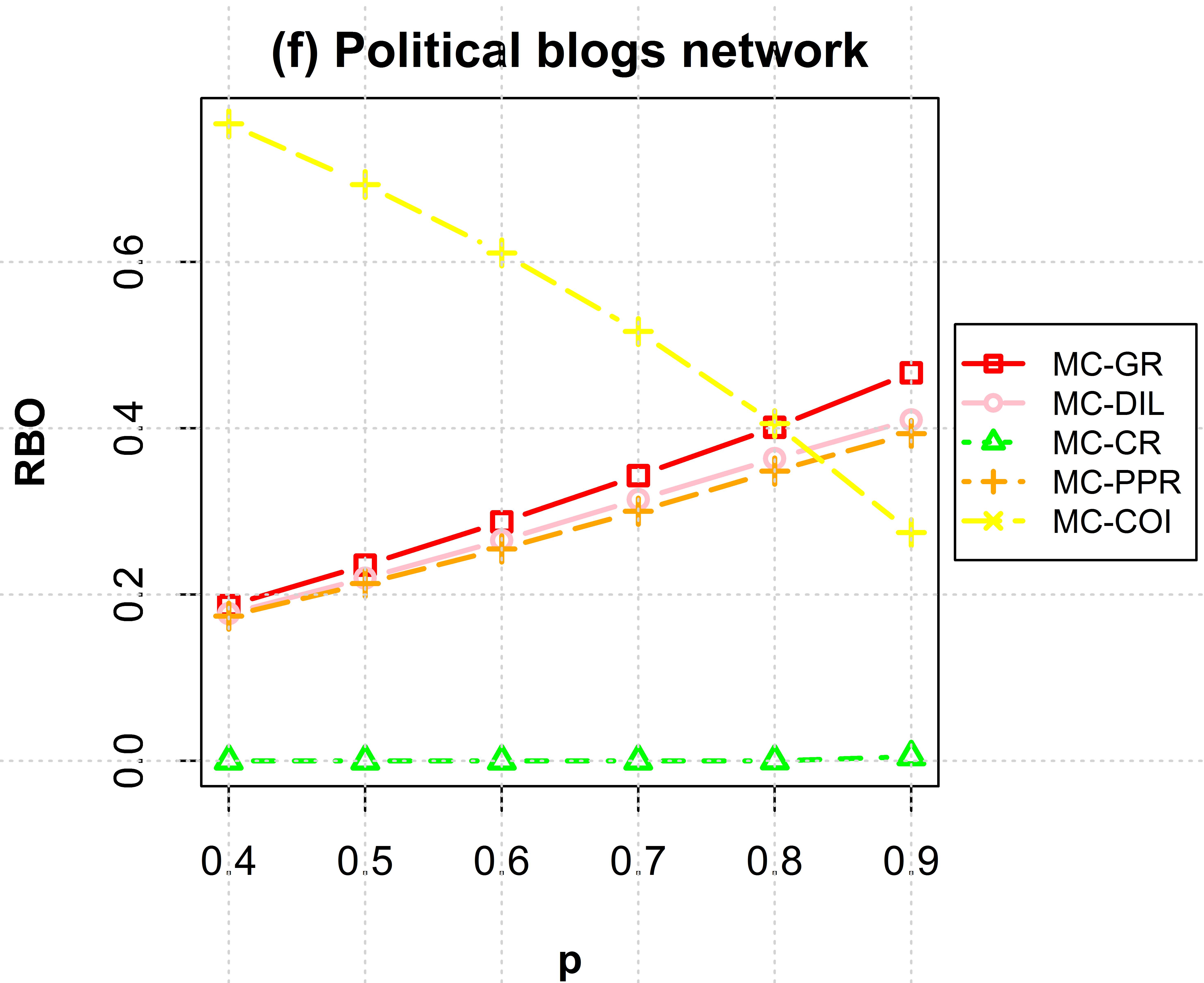}
\label{appfig:7}}
\hfil
\subfigure{
\includegraphics[width=2.3in]{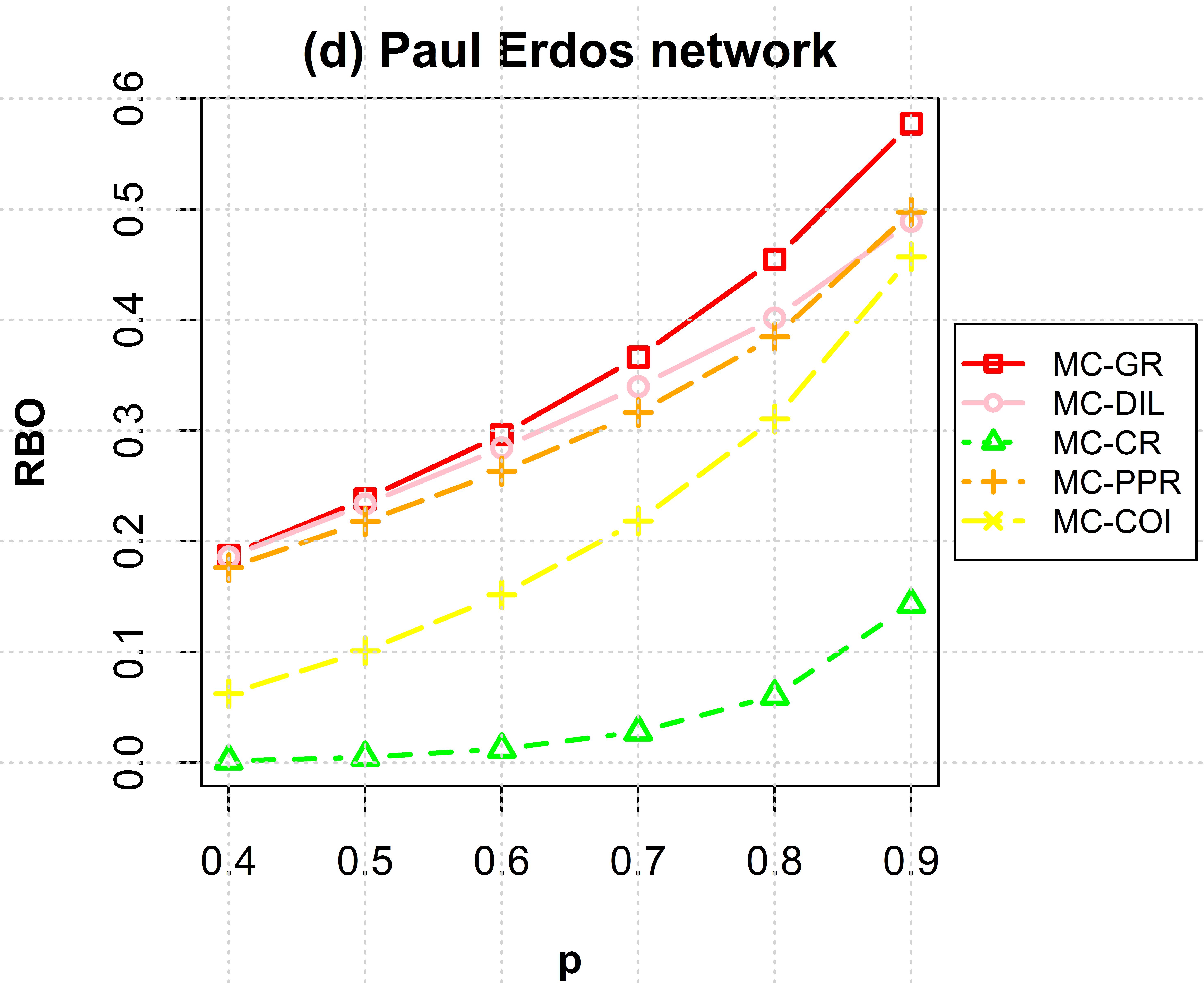}
\label{appfig:8}}
\hfil
\caption{The RBO values between the ranking lists produced by the different centralities
and the obtained ranking list from M-Centrality on the real-world networks under study.}
\label{appfigure1}
\end{figure*}

The results in Figure \ref{appfigure1} show that for Les Misérables, Game Of Thrones, Paul Erdős and E-mail networks, M-Centrality has higher concordance with Gravity centrality in high ranks. Note that in the cases of Game Of Thrones and E-mail networks, this correlation decreases as p increases. For Dolphins and US airport networks, M-Centrality achieves the highest concordance with Personalized PageRank, while for Netscience and Political blogs it is with Collective Influence for all values of p except for p=0.9 in Political blogs where the correlation in high ranks is registered with Gravity centrality. These results show that despite M-Centrality being highly correlated with Gravity centrality, the correlation in high ranks may vary between the two depending on the depth of the comparison.

\section{Correlation with the SIR model given transmission rate values}\label{regsol3}

\setcounter{figure}{0}
\renewcommand\thefigure{\Alph{section}.\arabic{figure}}
\setcounter{table}{0}
\renewcommand\thetable{\Alph{section}.\arabic{table}}

Table \ref{apptable5} show the numerical results corresponding to Figure \ref{figure 2} for the two most competitive measures, M-Centrality and Gravity. Although it has been shown previously that the two methods are highly correlated, one can see that in the majority of cases, the proposed method performs better than Gravity centrality. 

\begin{table}[h!]
\renewcommand{\arraystretch}{1}
\caption{Kendall tau correlation coefficient of M-Centrality and Gravity centrality with the SIR model. The number $20\%$ refers to $\beta = 0.2$$\times$$\beta_{th}$.}
\label{apptable5}
\centering
\resizebox{5in}{!}{
\begin{tabular}{|c|| c|  c  c c c c c c c|}
    \hline
    Network& $\beta$ & $20\%$ &    $40\%$ &     $60\%$  &     $80\%$ &     $100\%$ & $120\%$ &   $140\%$ &  $160\%$ \\
    \hline
    \multirow{2}{*}{Dolphins}               &$\tau(M, SIR)$  &     0.690 &  0.689 & 0.676& 0.760&  0.746& 0.763& 0.793& 0.877  \\
            &$\tau(Gr, SIR)$ &      0.661 &0.633 &0.667 &0.742 &0.723 &0.749 &0.779 &0.807               \\

  \hline\hline
  
\multirow{2}{*}{Les Misérables} &$\tau(M,SIR)$  &     0.690  & 0.737  & 0.748  &  0.874  &   0.852   &0.838  &   0.818 &   0.873               \\
&$\tau(Gr,SIR)$ &               0.717 &0.783 &0.801 &0.839 &0.802 &0.824 &0.801& 0.814       \\

  \hline\hline
 
\multirow{2}{*}{Game Of Thrones} &$\tau(M,SIR)$  &        0.719 &  0.736   & 0.792   &   0.798  &   0.816  & 0.846 &  0.879  &   0.861           \\
&$\tau(Gr,SIR)$ &                0.679 &0.726 &0.728& 0.748& 0.799 &0.831& 0.828 &0.821     \\

  \hline\hline
  
\multirow{2}{*}{Paul Erdős collaborations } &$\tau(M,SIR)$  &         0.746   & 0.783  &   0.798&    0.765&     0.850   &  0.845 &   0.850&    0.887 \\
&$\tau(Gr,SIR)$ &                 0.689& 0.749 &0.790& 0.789 &0.801& 0.804 &0.832 &0.840    \\

  \hline\hline
  
\multirow{2}{*}{Netscience}&$\tau(M,SIR)$  &         0.484&   0.486  & 0.485  & 0.487   &0.489 & 0.488&  0.487&   0.486 \\
&$\tau(Gr,SIR)$ &                  0.409& 0.419 &0.422 &0.428& 0.431 &0.430 &0.432 &0.433   \\

  \hline\hline

\multirow{2}{*}{US airport}&$\tau(M,SIR)$  &     0.393& 0.458 &0.485& 0.516& 0.540& 0.551& 0.544& 0.598             \\
&$\tau(Gr,SIR)$ &               0.378 &0.441& 0.470 &0.490& 0.506& 0.518& 0.535 &0.550         \\

  \hline\hline
  
\multirow{2}{*}{E-mail}&$\tau(M,SIR)$  &     0.410 &0.488& 0.523 &0.516& 0.575 &0.617& 0.605 &0.596                    \\
&$\tau(Gr,SIR)$ &            0.393 &0.462 &0.505& 0.512 &0.541 &0.523 &0.519 &0.554            \\

  \hline\hline
  
\multirow{2}{*}{US Political blogs}&$\tau(M,SIR)$  &    0.443& 0.538 &0.574 &0.643& 0.678 &0.608& 0.687& 0.745          \\
&$\tau(Gr,SIR)$ &       0.443 &0.518& 0.574& 0.597 &0.621 &0.611 &0.641 &0.647  \\

\hline
\end{tabular}
}
\end{table}

\section*{References}
\bibliographystyle{unsrt}
\bibliography{mybibf}

\end{document}